\documentclass{siamltex}

\usepackage{latexsym}
\usepackage{algorithm,algorithmic}
\newtheorem{property}{Property}
\newtheorem{claim}{Claim}

\def\mO{\mbox{O}}
\newcommand{\nop}[1]{}
\newcommand{\qed}{$\Box$}

\title{A linear time algorithm for \boldmath{$L(2,1)$}-labeling of trees\thanks{
A part of this article was presented in \emph{Proceedings of the 
Algorithms - ESA 2009, 17th Annual European Symposium, Copenhagen, Denmark, September 7-9, 2009}, Lecture Notes in Computer Science, Vol.~5757, pp.~35--46, Springer, 2009\cite{HIOU09}.}}

\author{Toru Hasunuma\thanks{
Department of Mathematical and Natural Sciences, The University of
Tokushima, Tokushima 770-8502, Japan. ({\tt hasunuma@ias.tokushima-u.ac.jp}).} 
\and Toshimasa Ishii\thanks{Department of Information and Management Science, Otaru University of Commerce, Otaru 047-8501, Japan. 
({\tt ishii@res.otaru-uc.ac.jp})} 
 \and  Hirotaka Ono\thanks{Department of Economic Engineering, Kyushu
University, Fukuoka 812-8581, Japan.
({\tt hirotaka@en.kyushu-u.ac.jp})
} \and Yushi Uno\thanks{Department of Mathematics and Information Sciences, Graduate School of Science, Osaka Prefecture University, Sakai 599-8531, Japan. ({\tt uno@mi.s.osakafu-u.ac.jp})}}

\begin{document}

\maketitle

\begin{abstract}
An $L(2,1)$-labeling of a graph $G$ is an assignment $f$ 
from the vertex set $V(G)$ to the set of nonnegative integers 
such that $|f(x)-f(y)|\ge 2$ if $x$ and $y$ are adjacent 
and $|f(x)-f(y)|\ge 1$ if $x$ and $y$ are at distance 2, 
for all $x$ and $y$ in $V(G)$. 
A $k$-$L(2,1)$-labeling is an $L(2,1)$-labeling $f:V(G)\rightarrow\{0,\ldots ,k\}$, 
and the $L(2,1)$-labeling problem asks the minimum $k$, 
which we denote by $\lambda(G)$, 
among all possible assignments. 
It is known that this problem is NP-hard even for
 graphs of treewidth 2, 
and tree is one of very few classes for which the
 problem is polynomially solvable. 
The running time of the best known algorithm for trees had been 
$\mO(\Delta^{4.5} n)$ for more than a decade, and 
an $\mO(\min\{n^{1.75},\Delta^{1.5}n\})$-time algorithm has appeared recently, 
where $\Delta$ and $n$ are the maximum degree and the number of vertices
 of an input tree, 
however, it has been open if it is solvable in linear time. 
In this paper, we finally settle this problem  
by establishing a linear time algorithm for $L(2,1)$-labeling of trees. 
Furthermore, we show that it can be extended to a linear time algorithm 
for $L(p,1)$-labeling with a constant $p$. 
\end{abstract}



\pagestyle{myheadings}
\thispagestyle{plain}
\markboth{T. HASUNUMA, T. ISHII, H. ONO AND Y. UNO}{
A LINEAR TIME ALGORITHM FOR $L(2,1)$-LABELING OF TREES}

\section{Introduction}
Let $G$ be an undirected graph. 
An {\em $L(p,q)$-labeling} of a graph $G$ is an assignment $f$ 
from the vertex set $V(G)$ to the set of nonnegative integers 
such that $|f(x)-f(y)|\ge p$ if $x$ and $y$ are adjacent 
and $|f(x)-f(y)|\ge q$ if $x$ and $y$ are at distance 2, 
for all $x$ and $y$ in $V(G)$. 
A {\rm $k$-$L(p,q)$-labeling} is an $L(p,q)$-labeling 
$f:V(G)\rightarrow\{0,\ldots ,k\}$, 
and the {\em $L(p,q)$-labeling problem} asks the minimum $k$ 
among all possible assignments. 
We call this invariant, the minimum value $k$, 
the {\em $L(p,q)$-labeling number} and 
denote it by $\lambda_{p,q}(G)$. 
Notice that we can use $k+1$ different labels when $\lambda_{p,q}(G)=k$ 
since we can use 0 as a label for conventional reasons. 

The original notion of $L(p,q)$-labeling can be seen 
in the context of frequency assignment, 
where `close' transmitters must receive frequencies 
that are at least $q$ frequencies apart 
and `very close' transmitters must receive frequencies 
that are at least $p$ $(\ge q)$ frequencies apart 
so that they can avoid interference. 

%
Among several possible settings of $p$ and $q$, $L(2,1)$-labeling
problem has been intensively and extensively studied due to its practical
importance.  
From the graph theoretical point of view, 
since this is a kind of vertex coloring problems, 
it has  attracted a lot of interest \cite{CK96,GY92,HRS08,W06}. 
%
We can find various related results on $L(2,1)$-labelings and
$L(p,q)$-labelings for other parameters in  comprehensive surveys by
Calamoneri \cite{C06} and by Yeh \cite{Y06}.  

\smallskip 

\noindent
{\bf Related Work:} 
There are also a number of studies on the $L(2,1)$-labeling problem, 
from the algorithmic point of view \cite{BKTL04,FKK01,KKL07}.   
It is known to be NP-hard for general graphs \cite{GY92}, 
and it still remains NP-hard for some restricted classes of graphs, 
such as planar graphs, bipartite graphs, chordal graphs \cite{BKTL04}, 
and it turned out to be NP-hard even for graphs of treewidth 2 
\cite{FGK05}.  
In contrast, only a few graph classes are known to have 
polynomial time algorithms for this problem, 
e.g., we can determine the $L(2,1)$-labeling number of 
paths, cycles, wheels within polynomial time \cite{GY92}.

As for trees, Griggs and Yeh \cite{GY92} showed that $\lambda(T)$ is
either $\Delta+1$ or $\Delta+2$ for any tree $T$, and also conjectured
that  determining  $\lambda(T)$
is
NP-hard, however, Chang and Kuo \cite{CK96} disproved this by presenting 
a polynomial time algorithm for computing $\lambda(T)$. 
Their algorithm exploits the fact that $\lambda(T)$
is either $\Delta+1$ or $\Delta+2$ for any tree $T$. 
Its running time is $\mbox{O}(\Delta^{4.5}n)$, where $\Delta$
is the maximum degree of a tree $T$ and $n=|V(T)|$.
%
%
%
%
This result has a great importance because it initiates to cultivate
polynomially solvable classes of graphs for the $L(2,1)$-labeling problem
and related problems. 
For example, Fiala et al. showed that it can be determined 
in $\mO(\lambda^{2t+4.5}n)$ time whether a $t$-almost tree has
$\lambda$-$L(2,1)$-labeling for $\lambda$ given as an input, where a
$t$-almost tree is a graph that can be a tree by 
eliminating $t$ edges \cite{FKK01}. Also, it was shown that the
$L(p,1)$-labeling problem for trees can be solved in
$\mO((p+\Delta)^{5.5}n) = \mO(\lambda^{5.5}n)$ 
time~\cite{CKKLY00}. Both results are based on Chang and Kuo's
algorithm, which is called as a subroutine in the algorithms. 
Moreover, the polynomially solvable result for 
trees holds for more general settings. The notion of $L(p,1)$-labeling
is generalized as $H(p,1)$-labeling, in which graph $H$ defines the
metric space of distances between two labels, whereas labels in
$L(p,1)$-labeling (that is, in $L(p,q)$-labeling) take nonnegative
integers; i.e., it is a special case that $H$ is a path graph.
In \cite{FGK08a}, it has been shown that the $H(p,1)$-labeling problem of
trees for arbitrary graph $H$ can be solved in polynomial time, 
which is also based on Chang and Kuo's idea. In passing, these
results are unfortunately not applicable for $L(p,q)$-labeling problems
for general $p$ and $q$. 
Recently, 
Fiala et al. ~\cite{FGK08b} showed that the $L(p,q)$-labeling problem 
for trees is NP-hard if $q$ is not a divisor of $p$, 
which is contrasting to the positive results mentioned above. 



As for $L(2,1)$-labeling of trees again, 
Chang and Kuo's $\mO(\Delta^{4.5}n)$ algorithm is the first 
polynomial time one. 
It is based on dynamic programming (DP) approach, 
and it checks whether $(\Delta+1)$-$L(2,1)$-labeling is possible or not 
from leaf vertices to a root vertex in the original tree structure. 
The principle of optimality requires to solve 
at each vertex of the tree the assignments of labels to subtrees, 
and the assignments are formulated as the maximum  matching 
in a certain bipartite graph. 
Recently, an $\mO(\min\{n^{1.75},\Delta^{1.5}n\})$ time algorithm 
has been proposed \cite{HIOU08}. 
It is based on the similar DP framework to Chang and Kuo's algorithm, 
but achieves its efficiency by reducing heavy computation 
of bipartite matching in Chang and Kuo's 
and by using an amortized analysis. 
We give a concise review of these two algorithms 
in Subsection \ref{CK-algo-sec}.

\nop{
As for $L(2,1)$-labeling of trees again, recently, an
$\mO(\min\{n^{1.75},\Delta^{1.5}n\})$ time algorithm has been proposed
\cite{HIOU08},  which substantially improves the previous result. 
Both Chang and Kuo's algorithm and the new algorithm are based on
dynamic programming (DP) approach, 
which checks whether $(\Delta+1)$-$L(2,1)$-labeling is possible or not 
from leaf vertices to a root vertex in the original tree
structure, and the principle of 
optimality requires to solve at each vertex of the tree the assignments
of labels to subtrees. The assignments are
formulated as the maximum  matching in a bipartite graph with $\mO(\Delta)$ vertices
and $\mO(\Delta^2)$ edges, that is, it takes $\mO(\Delta^{2.5})$ time \cite{HK73}. 
Since the assignment at each vertex should be solved $\Delta^2$
times to fill the DP table up, 
the total running time of Chang and Kuo's algorithm is
$\mO(\Delta^{4.5}n)$. The $\mO(n^{1.75})$-time algorithm in
\cite{HIOU08} circumvents
this heavy computation by two ways: one is  that it is
sufficient to solve the bipartite matching not $\Delta^2$ times 
but essentially 
$\Delta$ times at each vertex;  
the other is  an amortized analysis, in which it can be shown that the 
bipartite matching problems should be solved roughly only
$\mO(n/\Delta)$ times in average with respect to degrees. Note that the
observation is derived from the fact 
that the assignments near leaf vertices can be solved
efficiently. 
We give a concise review of these two algorithms 
in Subsections \ref{CK-algo-sec} and \ref{sec:n175}, respectively. 
}

\smallskip 

\noindent
{\bf Our Contributions:} 
Although there have been a few polynomial time algorithms 
for $L(2,1)$-labeling of trees, 
it has been open if it can be improved to linear time~\cite{C06}. 
In this paper, we present a linear time algorithm for $L(2,1)$-labeling
of trees, which finally settles this problem. 
The linear time algorithm is based on DP approach, which is adopted in 
the preceding two polynomial time algorithms~\cite{CK96,HIOU08}. 
Besides using their ideas, we newly introduce the notion 
of ``label compatibility'', which indicates how we flexibly change
labels with preserving its
$(\Delta+1)$-$L(2,1)$-labeling. 
The label compatibility is a quite handy notion and in fact, it deduces 
several useful facts on $L(2,1)$-labeling of trees in fact. For
example, it is used to show that only 
$\mO(\log_{\Delta} n)$ labels are essential for $L(2,1)$-labeling in any
input tree, which enables to replace the bipartite matching of graphs
with the maximum flow of much smaller networks as an engine 
for filling out DP tables.  
Roughly speaking, by utilizing the label compatibility in several 
ways, we can obtain a linear time 
algorithm for $L(2,1)$-labeling of trees. 
It should be noted that the notion of label compatibility
can be generalized for $L(p,1)$-labeling of trees, and therefore, 
we can obtain a linear time algorithm for $L(p,1)$-labeling of trees 
with a constant $p$ by extending the one for $L(2,1)$-labeling 
together with the generalized label compatibility. 



\medskip 

\noindent
{\bf Organization of this Paper:}  
Sections from \ref{section-preliminaries} to
\ref{section-linear-running-time} deal with 
$L(2,1)$-labeling of trees, and Section \ref{sec:lp1} deals with its
extension to $L(p,1)$-labeling of trees. 
Section \ref{section-preliminaries} gives basic definitions 
and introduces as a warm-up 
the ideas of Chang and Kuo's $\mO(\Delta^{4.5}n)$ time algorithm 
and its improvement into $\mO(n^{1.75})$ time. 
Section \ref{section-label-compatibility} 
introduces the crucial notion of label compatibility that can bundle a
set of compatible vertices and reduce the size of the graph   
constructed for computing bipartite matchings. 
Moreover, this allows to use maximum-flow based computation for them. 
In Section \ref{section-linear-running-time}, 
we give precise analyses to achieve linear running time for
$L(2,1)$-labeling of trees. 
In Section \ref{sec:lp1}, 
after generalizing 
the techniques introduced in Section \ref{section-label-compatibility} 
to $L(p,1)$-labeling, 
we show that analyses similar to those in Section
\ref{section-linear-running-time} 
are possible for a constant $p$; 
$L(p,1)$-labeling of trees with a constant $p$ is linearly solvable also. 
Section \ref{sec:proof-level-lem} provides a proof of the key lemma used
in Sections \ref{section-label-compatibility} - \ref{sec:lp1}, named
Level Lemma, and Section \ref{sec:conclusion} concludes the paper.
\section{Preliminaries}
\label{section-preliminaries}

\subsection{Definitions and Notations}
A graph $G$ is an ordered pair of its vertex set $V(G)$ and edge set $E(G)$ 
and is denoted by $G=(V(G),E(G))$. 
We assume throughout this paper that all graphs are undirected, 
simple and connected, unless otherwise stated. 
Therefore, an edge $e\in E(G)$ is an unordered pair of vertices $u$ and $v$, 
which are {\em end vertices} of $e$, 
and we often denote it by 
$e=(u,v)$.
Two vertices $u$ and $v$ are {\em adjacent} if 
$(u,v)\in E(G)$.
For a graph $G$, we may denote $|V(G)|$, 
the number of vertices of $G$, simply by $|G|$. 
A graph $G=(V(G),E(G))$ is called {\em bipartite} if the vertex set
$V(G)$ can be divided into two disjoint sets $V_1$ and $V_2$ such that
every edge in $E(G)$ connects a vertex in $V_1$ and one in $V_2$; such
$G$ is denoted by $(V_1,V_2,E)$. 
Throughout the paper, we also denote by $n$ the number of vertices
of input tree $T$. 

For a graph $G$, the ({\em open}) {\em neighborhood} of a vertex $v\in V(G)$ 
is the set $N_G(v)=\{u\in V(G)\mid (u,v)\in E(G)\}$, 
and the {\em closed neighborhood} of $v$ 
is the set $N_G[v]=N_G(v)\cup\{v\}$. 
The {\em degree} of a vertex $v$ is $|N_G(v)|$, 
and is denoted by $d_G(v)$. 
We use $\Delta(G)$ to denote the 
maximum degree of a graph $G$. 
A vertex whose degree is $\Delta(G)$ is called {\em major}. 
We often drop $G$ in these notations if there are no confusions. 
A vertex whose degree is $1$ is called a {\em leaf vertex}, or simply a
{\em leaf}. 
When we describe algorithms, 
it is convenient
to regard the input tree to be rooted at
a leaf vertex $r$. 
Then we can define the parent-child relationship on vertices 
in the usual way. 
For a rooted tree, its {\it height} is the length of the longest path 
from the root to a leaf. 
For any vertex $v$, 
the set of its children 
is denoted 
by $C(v)$. 
For a vertex $v$, define $d'(v)=|C(v)|$.

\subsection{Chang and Kuo's Algorithm and its Improvement}
\label{CK-algo-sec}
\label{sec:n175}

Before explaining algorithms, we give 
some significant properties on $L(2,1)$-labeling 
of graphs or trees 
that have been used so far for designing $L(2,1)$-labeling algorithms. 
We can see that $\lambda(G)\ge \Delta+1$ holds for any graph $G$.  
Griggs and Yeh \cite{GY92} 
observed that any major vertex in $G$ must be labeled $0$ or $\Delta+1$
when $\lambda(G) = \Delta+1$, and that 
if $\lambda(G) = \Delta+1$, 
then $N_{G}[v]$ contains at most two major vertices for any $v \in V(G)$.
Furthermore, they showed that $\lambda(T)$ is either $\Delta+1$ or $\Delta+2$ 
for any tree $T$. 
\nop{Concerning this fact, they also conjectured that 
determining if $\lambda(T)$ is $\Delta+1$ or $\Delta+2$ is NP-hard, 
however, Chang and Kuo \cite{CK96} disproved this by presenting 
an $\mO(\Delta^{4.5}n)$ time algorithm for computing $\lambda(T)$. 
}
By using this fact, Chang and Kuo \cite{CK96} presented
 an $\mO(\Delta^{4.5}n)$ time
algorithm for computing $\lambda(T)$.  



\medskip
\noindent
{\bf Chang and Kuo's Algorithm}\ \ 
Now, we first review the idea of Chang and Kuo's dynamic programming algorithm 
({\sc CK} algorithm)
for the $L(2,1)$-labeling problem of trees, 
since our linear time algorithm also depends on the same formula of 
the principle of optimality. 
The algorithm determines if $\lambda(T) = \Delta + 1$, and if so, 
we can easily construct the labeling with $\lambda (T) = \Delta + 1$.

To describe the idea, we introduce some notations. 
We assume for explanation that $T$ is rooted at some leaf vertex $r$. 
Given a vertex $v$, we denote the subtree of $T$ rooted at $v$ by 
$T(v)$. Let $T(u,v)$ be a tree rooted at $u$ that forms
$T(u,v)=(\{u\}\cup V(T(v)),\{(u,v)\}\cup E(T(v)))$. Note that this $u$
is just a virtual vertex for explanation and $T(u,v)$ is
uniquely determined by $T(v)$. 
For $T(u,v)$, we define 
\begin{eqnarray*}
\delta((u,v),(a,b)) & = & \left\{
 \begin{array}{ll}
  1, & \mbox{ if } \  \lambda(T(u,v) \mid f(u)=a, f(v)=b)\le \Delta + 1, \\
  0, & \mbox{ otherwise,}
 \end{array}
\right. 
\end{eqnarray*}
where $\lambda(T(u,v) \mid f(u)=a, f(v)=b)$ denotes 
the minimum $k$ of a $k$-$L(2,1)$-labeling $f$ on $T(u,v)$ satisfying
$f(u)=a$ and $f(v)=b$. 
This $\delta$ function satisfies the following formula: 
\begin{eqnarray*}
 \delta((u,v),(a,b)) &\! =\!& \left\{
 \begin{array}{ll}
  1, & \!\mbox{ if there is an injective assignment } g\!: 
   C(v) \rightarrow \{0,1, \ldots, \Delta\!+\!1\}\!\\
  &  -\{a,b-1,b,b+1\} 
   \mbox{ such that }\delta((v,w),(b,g(w)))=1\mbox{ for
   each }\\ 
  & w \in C(v),\\
  0, & \mbox{ otherwise. }
  

 \end{array}
\right.
\end{eqnarray*}
The existence of such an injective assignment $g$ 
is formalized as the maximum 
matching problem: For a bipartite graph
$G(u,v,a,b)=(C(v),X, 
E(u,v,a,b))$, where 
$X=\{0,1,$ $\ldots,\Delta,\Delta+1\}$  and 
$E(u,v,a,b)=\{(w,c) \mid \delta((v,w),(b,c))=1, c\in X-
\{a\}, w\in C(v)\}$, 
we can see that there is an injective assignment $g$:
$C(v) \rightarrow \{0,1,\ldots,\Delta+1\}-\{a,b-1,b,b+1\}$
if there exists a matching of size $d'(v)$ in $G(u,v,a,b)$. 
Namely, for $T(u,v)$ and two labels $a$ and $b$, we can easily (i.e., in
polynomial time) determine the value of $\delta((u,v),(a,b))$ if the
values of $\delta$ function for $T(v,w), w \in C(v)$ and any two pairs of labels
are given. 
Now let $t(v)$ be the time for calculating $\delta((u,v),(*,*))$ 
for vertex $v$, where $\delta((u,v),(*,*))$ denotes 
$\delta((u,v),(a,b))$ for all possible pairs of $a$ and $b$ 
under consideration. (We use $*$ also in 
some other places in a similar manner.) 
{\sc CK} algorithm solves this bipartite matching
problems of $\mO(\Delta)$ vertices and $\mO(\Delta^2)$ edges
$\mO(\Delta^2)$ times for each $v$, in order to obtain $\delta$-values
for all pairs of labels $a$ and $b$. 
This amounts $t(v)=\mO(\Delta^{2.5})\times \mO(\Delta^2)=\mO(\Delta^{4.5})$, 
where the first $\mO(\Delta^{2.5})$ is the time complexity 
of the bipartite matching problem \cite{HK73}. 
Thus the total running time
is $\sum_{v\in V}t(v)=\mO(\Delta^{4.5}n)$.

\medskip
\noindent
{\bf An \boldmath{$\mO(n^{1.75})$}-time Algorithm}\ \ 
Next, we review the 
$\mO(n^{1.75})$-time algorithm 
proposed in \cite{HIOU08}.  
The running time $\mO(n^{1.75})$ is roughly achieved by two strategies. 
One is that the problem can be solved by a simple linear time
algorithm if $\Delta=\Omega(\sqrt{n})$, 
and the other is that it can be solved in $\mO(\Delta^{1.5}n)$ time 
for any input tree. 
\nop{
In this subsection, we review the ideas of the $\mO(n^{1.75})$-time algorithm 
presented in \cite{HIOU08}, which  is
  described in Algorithm
\ref{alg:label-tree} in Appendix.
The running time $\mO(n^{1.75})$ is roughly achieved by two strategies. 
One is that the problem can be solved by a simple linear time
algorithm if $\Delta=\Omega(\sqrt{n})$. 
The other is that we can solve the problem for any input tree in
$\mO(\Delta^{1.5}n)$ time. 
}

The first idea of the speedup is that for computing $\delta((u,v),(*,b))$, 
the algorithm does not solve the bipartite matching problems 
every time from scratch, but reuse the obtained matching structure.
More precisely, 
the bipartite matching problem is solved 
for $G(u,v,-,b)=(C(v),X,E(u,v,-,b))$ instead of $G(u,v,a,b)$ 
for a specific $a$,
where $E(u,v,-,b)=\{(w,c) \mid \delta((v,w),(b,c))=1, c\in X, w\in C(v)\}$. 
A maximum matching of 
$G(u,v,-,b)$ is observed to satisfy the following properties: 
\nop{
If $\delta((u,v),(i,b)) = 1$ holds for some label $i$, then
$G(u,v,-,b)$ has a matching $M$ of size $d'(v)$ such that vertex
$i\in X$ is not matched by $M$. Suppose that $G(u,v,-,b)$ has a matching
$M$ of size $d'(v)$. 
Clearly, an unmatched vertex $i \in X$ satisfies $\delta((u,v),(i,b)) =
1$. For a matched vertex $i' \in X$, if there is an $M$-alternating path
from some unmatched vertex to vertex $i'$, then we can obtain another
matching $M'$ of size $d'(v)$ such that $i'$ is not matched by $M'$,
i.e., $\delta((u,v),(i',b)) = 1$. Thus, the following properties hold.
}
\begin{property}\label{alternate1-prop}
If $G(u,v,-,b)$ has no matching of size $d'(v)$,  
then 
$\delta((u,v),$ $(i,b))$ $=0$ for any label $i$.
\end{property}
\begin{property}\label{alternate-prop}
$\delta((u,v),(i,b))=1$ if and only if
vertex $i$ can be reached by an $M$-alternating path
from some vertex in $X$ unmatched by $M$ in $G(u,v,-,b)$,
where $M$ denotes a maximum matching of $G(u,v,-,b)$ $($of size
 $d'(v))$.
\end{property}

From these properties, 
 $\delta((u,v),(*,b))$ can be computed by
a single bipartite matching  and a single graph search, and
its total running time  is
$\mO(\Delta^{1.5}d'(v))+\mO(\Delta d'(v))=\mO(\Delta^{1.5}d'(v))$ 
(for solving the bipartite matching of $G(u,v,-,b)$, which has
$\mO(\Delta)$ vertices and $\mO(\Delta d'(v))$ edges, and for 
a single graph search). Since this calculation is done for all $b$, we
have $t(v)=\mO(\Delta^{2.5}d'(v))$.

The other technique of the speedup introduced in \cite{HIOU08} 
is based on preprocessing operations
for amortized analysis.
By some preprocessing operations, the shape of input trees 
can be restricted 
while preserving $L(2,1)$-labeling number, 
and the input trees can be assumed to 
satisfy the following two properties.
%
%
\begin{property}\label{leaf-prop}
 All vertices adjacent to a leaf vertex are major vertices. 
\end{property}
\begin{property}\label{P-prop}
 The size of any path component of $T$ is at most $3$.
\end{property}
%

Here, a sequence of vertices $v_1,v_2,\ldots ,v_{\ell}$ is called
a {\em path component} if $(v_i,v_{i+1})\in E$ for all $i=1,2,\ldots ,\ell-1$ 
and $d(v_i)=2$ for all $i=1,2,\ldots ,\ell$, 
and  $\ell$ is called the {\em size} of the path component.

Furthermore, this preprocessing operations enable the following amortized analysis.
Let $V_L$ and $V_Q$ be the set of leaf vertices and the set of major vertices
whose children are all leaf vertices, respectively.
Also, let $d''(v) = |C(v)-V_L|$ for $v \in V$.
(Note that $d''(v) = 0$ for $v \in V_L \cup V_Q$.)

By Property~\ref{leaf-prop}, if we go down the resulting tree from 
a root, then we reach a major vertex in $V_Q$. 
Then, the following facts are observed: (i) 
for $v\in V_Q$ $\delta((u,v),(a,b))=1$ if and only if $b=0$
 or
$\Delta+1$ and $|a-b|\ge 2$, (ii) $|V_Q|\le n/\Delta$. 
Note that (i) implies that it is not required to solve the bipartite
matching to obtain $\delta$-values. Also (ii) and Property \ref{P-prop}
imply that $|V-V_Q-V_L|=\mO(n/\Delta)$ 
 (this can be obtained by pruning leaf vertices and
regarding $V_Q$ vertices as new leaves).
Since it is not necessary to compute bipartite
matchings for $v\in V_L\cup V_Q$, 
and this implies that 
the total time to obtain $\delta$-values for all $v$'s is 
$\sum_{v\in V}t(v)=\mO(\sum_{v\in V-V_L-V_Q}t(v))$, 
which turned out to be $\mO(\Delta^{2.5}\sum_{v\in V-V_L-V_Q}d''(v))$.
Since $\sum_{v\in V-V_L-V_Q}d''(v)=|V-V_L-V_Q|+|V_Q|-1=\mO(n/\Delta)$,
we obtain $\sum_{v\in V-V_L-V_Q}t(v)=\mO(\Delta^{1.5}n)$. 
Since we have a linear time algorithm if 
$\Delta=\Omega(\sqrt{n})$ as mentioned above, we can solve the
problem in $\mO(n^{1.75})$ time in total.

\subsection{Concrete Form of \boldmath{$\mO(n^{1.75})$}-time Algorithm}
We describe 
the main routine of $\mO(n^{1.75})$-time algorithm \cite{HIOU08} 
as Algorithm \ref{alg:label-tree}. This algorithm calls two  
algorithms as subroutines. 
One is Algorithm \ref{alg:update-matching}, which is used to quickly compute 
$\delta((u,v),(*,b))$, as explained in Section \ref{sec:n175}. 
The other is Algorithm \ref{alg:preprocessing}, which does  
the preprocessing operations for an input tree. 

 They are carried out 
(i) to remove the vertices that are `irrelevant' to the
$L(2,1)$-labeling number, 
and (ii) to divide $T$ into several subtrees that preserve the
$L(2,1)$-labeling number. It is easy to show that neither of the operations affects the $L(2,1)$-labeling number. 
Note that, these operations may not 
reduce the size of the input tree,
but more importantly, they restrict the shape of input trees, which
enables an amortized analysis.

We can verify that after the preprocessing operations, the input trees satisfy 
Properties \ref{leaf-prop} and \ref{P-prop}.

\begin{algorithm}[htb]
 \caption{{\sc Label-Tree}}\label{alg:label-tree}
 \begin{algorithmic}[1]
\STATE  {\small  Do {\sc Preprocessing} (Algorithm \ref{alg:preprocessing}). 
  \STATE If $N[v]$ contains at least three major vertices for some vertex
  $v \in V$, output ``No''. Halt.
  \STATE If the number of major vertices is at most $\Delta-6$, 
output ``Yes''. 
   Halt.
  \STATE For $T(u,v)$ with $v \in V_Q$ (its height is 2), let 
  $\delta((u,v),(a,0)):=1$ for each label $a \neq 0,1$,  
  $\delta((u,v),(a,\Delta+1)):=1$ for each label $a \neq \Delta,\Delta+1$, 
  and $\delta((u,v),(*,*)):=0$ for any 
  other pair of labels. Let $h:=3$.  
  \STATE   For all $T(u,v)$ of height $h$, compute
  $\delta((u,v),(*,*))$ (Compute $\delta((u,v),(*,b))$ for each $b$ by
  {\bf {\sc Maintain-matching}$(G(u,v,-,b))$} (Algorithm
  \ref{alg:update-matching}). 
  \STATE  If $h=h^*$ where $h^*$ is the height of root $r$ of $T$, then
  goto Step 7. Otherwise let $h:=h+1$ and goto Step 4. 
  \STATE If $\delta((r,v),(a,b))=1$ for some $(a,b)$, then 
  output ``Yes''; otherwise output ``No''. Halt. }
 \end{algorithmic}
\end{algorithm}

 \begin{algorithm}
  \caption{{\sc Maintain-Matching}($G(u,v,-,b)$)}\label{alg:update-matching}
  \begin{algorithmic}[1]
   \STATE {\small 
 Find a maximum bipartite matching $M$ of $G(u,v,-,b)$. If
   $G(u,v,-,b)$ has no matching of size $d'(v)$, output
   $\delta((u,v),(*,b))$ as $\delta((u,v),(i,b))=0$ for every label $i$.
   \STATE Let $X'$ be the set of unmatched vertices under $M$.
   For each label vertex $i$ that is reachable from a vertex in $X'$ via
   $M$-alternating path, let $\delta((u,v),(i,b))=1$. For the other
   vertices $j$, let $\delta((u,v),(j,b))=0$. Output
   $\delta((u,v),(*,b))$.} 
  \end{algorithmic}
  \end{algorithm}

\begin{algorithm}
 \caption{{\sc Preprocessing}}\label{alg:preprocessing}
\begin{algorithmic}[1]
 \STATE {\small Check if there is a leaf $v$ whose unique neighbor $u$
	    has degree less than $\Delta$. If so, remove $v$ and edge
	    $(u,v)$ from $T$ until such a leaf does not exist.
 \STATE  Check if there is a path component whose size is at least
	    $4$, say $v_1, v_2, \ldots, v_{\ell}$, and let $v_0$ and  
	    $v_{\ell +1}$ be the unique adjacent vertices of $v_1$ and
	    $v_{\ell}$  other than $v_2$ and $v_{\ell -1}$,
	    respectively.  
	    If it exists, assume $T$ is rooted at $v_1$, divide $T$ into $T_1:=T(v_1,v_0)$ and
	    $T_2:=T(v_{4},v_{5})$, and remove
	    $v_2$ and $v_3$. Continue this operation until 
	    such a path 
	    component does not exist.}            
\end{algorithmic}
 \end{algorithm}

\section{Label Compatibility and Flow-based Computation of \boldmath{$\delta$}}
\label{section-label-compatibility}
\nop{
As mentioned in Subsection \ref{sec:n175}, we can achieve an efficient
computation of $\delta$-values by reusing the matching structures, which
is one of keys of the running time $\mO(n^{1.75})$. 
} 
As reviewed in Subsection \ref{sec:n175}, 
one of keys of an efficient computation of $\delta$-values 
was reusing the matching structures. 
In this section, for a further speedup of the computation of $\delta$-values, 
we introduce a new novel notion, which we call `label compatibility', 
that enables to treat several labels equivalently 
under the computation of $\delta$-values. 
Then, the faster computation of $\delta$-values is achieved 
on a maximum flow algorithm instead of a maximum matching algorithm.  
Seemingly, this sounds a bit strange,
because the time complexity of the maximum flow problem is greater than the one 
of the bipartite matching problem.
The trick is that by this notion the new flow-based computation 
can use a smaller network (graph) 
than the graph $G(u,v,-,b)$ used in the bipartite matching. 

\subsection{Label Compatibility and Neck/Head Levels}
\label{sec:label-compatibility}
Let $L_h=\{h,h+1,\ldots,$ $\Delta-h, \Delta-h+1\}$. 
Let $T$ be a tree rooted at $v$, and $u\not\in V(T)$. 
We say that $T$ is {\em head}-$L_h$-{\em compatible} if $\delta((u,v),(a,b))=\delta((u,v),(a',b))$ for all $a,a'\in L_h$ and $b\in L_0$ with
             $|a-b|\ge 2$ and $|a'-b|\ge 2$. Analogously, we say that
             $T$ is {\rm neck}-$L_h$-{\rm compatible} if
             $\delta((u,v),(a,b)) = \delta((u,v),(a,b'))$ for all $a\in
             L_0$ and $b,b' \in L_h$
             with $|a-b|\ge 2$ and $|a-b'|\ge 2$. 
The neck and head levels of $T$ are defined as follows:

\medskip

\begin{definition}\label{def:labelcomatibility21}
 Let $T$ be a tree rooted at $v$, and $u\not\in V(T)$.\\
{\rm (i)}       The {\rm neck level} $($resp., {\rm head level}$)$ of $T$ is $0$ if $T$ is 
             neck-$L_0$-compatible $($resp., head-$L_0$-compatible$)$. 
{\rm (ii)}
             The {\rm neck level} $($resp., {\rm head level}$)$ of $T$ is
             $h\ (\geq 1)$ if $T$ is not
             neck-$L_{h-1}$-compatible $($resp., head-$L_{h-1}$-compatible$)$
             but neck-$L_h$-compatible $($resp., head-$L_h$-compatible$)$. 
\end{definition}

\medskip

\noindent
An intuitive explanation of neck-$L_h$-compatibility (resp.,
head-$L_h$-compatibility) of $T$ is that 
if for $T(u,v)$, 
a label in
$L_h$ is assigned to
$v$ (resp., $u$)   under $(\Delta+1)$-$L(2,1)$-labeling of $T(u,v)$,  the label can
be replaced with another label in $L_{h}$ without violating a proper
$(\Delta+1)$-$L(2,1)$-labeling; 
labels in $L_h$ are compatible. The neck and head levels of $T$
represent the bounds of $L_h$-compatibility of $T$. 
Thus, a trivial bound on neck and head levels is $(\Delta+1)/2$.

For the relationship between the neck/head 
levels and the tree size, we can show the following lemma and theorem 
that are crucial throughout analyses for the linear time algorithm. 

\medskip

\begin{lemma}\label{level-lem}(Level Lemma)\rm 
Let $T'$ be a subtree of $T$.
If $|T'|< (\Delta-3-2h)^{h/2}$ and $\Delta-2h\geq 2$, then
the head level and neck level of $T'$ are both at most $h$. 
\end{lemma}

\medskip

\begin{theorem}\rm\label{theo:logn}
 For a tree $T$, both the
 head and neck levels of $T$ are bounded by $\mO(\log |T|/\log \Delta)$.
\end{theorem}

\medskip 

We give proofs of Lemma \ref{level-lem} (Level Lemma) and Theorem
\ref{theo:logn} in Section \ref{sec:proof-level-lem} as those in more
generalized forms 
(see Lemma \ref{level-lem2} and Theorem \ref{theo:logn2}). 

\subsection{Flow-based Computation of \boldmath{$\delta$}}\label{maxflow-sec}
We are ready to explain the faster computation of $\delta$-values. 
Recall that $\delta((u,v),(a,b))$ $=1$ holds if there exists a matching of
$G(u,v,a,b)$ in which all $C(v)$ vertices are just matched; which
vertex  is matched to a vertex in $X$ does not matter. {From} this
fact, we can treat vertices in $X$ corresponding to 
$L_h$
equally in
computing $\delta$, if $T$ is neck- and head-$L_h$-compatible. 
The idea of the fast computation of $\delta$-values is that,
by bundling compatible vertices in $X$ of $G$, we reduce the size of
a graph (or a network) for computing the assignments of labels,  which is no
longer the maximum matching; the maximum flow. 

\begin{figure*}[ht]
  \begin{center}
   \input{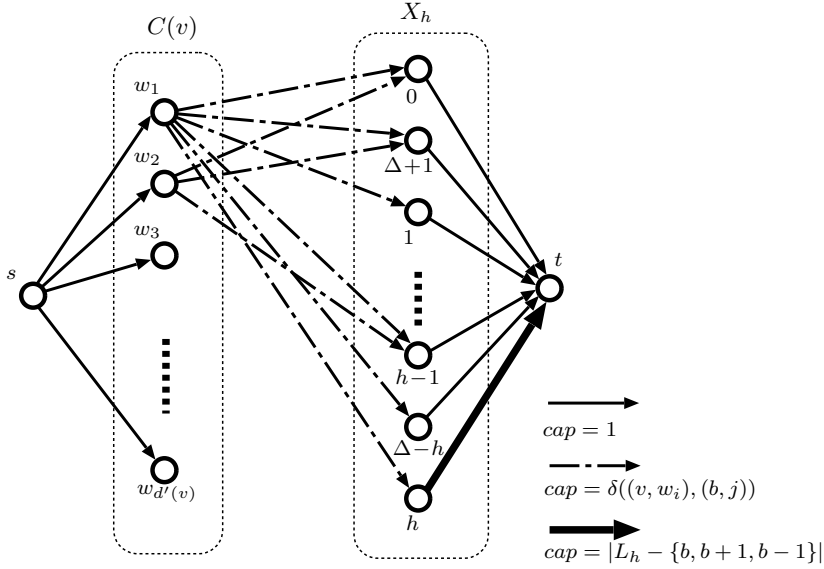}
  \end{center}
 \caption{An example of ${\cal N}(u,v,-,b)$ where $h=h(v)$.}
 \label{fig:flow}
\end{figure*}

The algorithm reviewed 
in Subsection \ref{sec:n175}
 computes
$\delta$-values not by solving the maximum matchings of $G(u,v,a,b)$ for
all pairs of $a$ and $b$ but by finding a maximum matching $M$ of $G(u,v,-,b)$
once and then searching $M$-alternating paths. 
In the new flow-based computation, we adopt the same strategy; 
for a tree $T(v)$ whose head and neck levels are at most $h(v)$, 
we do not prepare a network for a specific pair 
$(a,b)$, say ${\cal N}(u,v,a,b)$, but a general network ${\cal
N}(u,v,-,b) = (\{s,t\}\cup C(v) \cup X_{h(v)},E(v)\cup E_X \cup E_{\delta},
cap)$, where 
 $X_{h(v)} =  (L_0-L_{h(v)}) \cup \{h(v)\}$, 
$E(v) = \{(s,w)\mid w\in
C(v)\}$, $E_X=\{(c,t) \mid c \in X_{h(v)}\}$, $E_{\delta}=\{(w,c) \mid w\in
C(v), c\in X_{h(v)}\}$, and $cap(e)$ function is defined as follows: 
$\forall e\in E(v)$, $cap(e)=1$, for $e=(w,c)\in E_{\delta}$, $cap(e)=1$
if $\delta((v,w),(b,c))=1$, $cap(e)=0$ otherwise, and for $e=(c,t)\in
E_X$, $cap(e)=1$ if $c \neq h(v)$, $cap(e)=|L_{h(v)}-\{b,b+1,b-1\}|$ if
$c=h(v)$. 
See Figure \ref{fig:flow} as an example of ${\cal N}(u,v,-,b)$. 

For a maximum flow $\psi : e \rightarrow R^+$, we define $X'$ as
$\{c\in X_{h(v)} \mid cap((c,t))-\psi((c,t))\ge 1 \}$. 
By the flow integrality and arguments similarly to
Properties \ref{alternate1-prop} and \ref{alternate-prop}, we can obtain the following properties:
\medskip

\begin{lemma}\rm
If ${\cal N}(u,v,-,b)$ has no flow of size $d'(v)$,  
then 
$\delta((u,v),(i,b))$ $=0$ for any label $i$.
\end{lemma}
\begin{lemma}
For $i\in L_0-L_{h(v)}$, $\delta((u,v),(i,b))=1$ if and only if 
vertex $i$ can be reached by a $\psi$-alternating path
from some vertex in $X'$ in ${\cal N}(u,v,-,b)$. 
For $i\in L_{h(v)}$, $\delta((u,v),(i,b))=1$ if and only if 
vertex $h(v)$ can be reached by a $\psi$-alternating path
from some vertex in $X'$ in ${\cal N}(u,v,-,b)$. 
\end{lemma}

\medskip

Here, a $\psi$-alternating path is defined as follows: Given a flow
$\psi$, a path in $E_\delta$
is called {\em $\psi$-alternating} if its edges alternately
satisfy $cap(e)-\psi(e)\ge 1$ and $\psi(e)\ge 1$.
By these lemmas, we can obtain $\delta((u,v),(*,b))$-values for $b$ by
solving the maximum flow of ${\cal N}(u,v,-,b)$ once and then applying a
single graph search. 

The current fastest algorithm for the maximum flow problem 
runs in $\mO(\min\{m^{1/2},$ $n^{2/3}\} m\log(n^2/m)$ $\log
U)=\mO(n^{2/3}m\log n\log U)$ time, where $U$, $n$ and $m$ 
are the maximum capacity of edges, the number of vertices and
edges, respectively~\cite{GR98}. Thus the running time of calculating
$\delta((u,v),(a,b))$ for a pair $(a,b)$ is
\[
 \mO((h(v)+d''(v))^{2/3}(h(v)d''(v))\log
(h(v)+d''(v))\log \Delta)
= \mO(\Delta^{2/3}(h(v)d''(v))\log^2 \Delta),  
\]
since $h(v)\le \Delta$ and
$d''(v)\le \Delta$ (recall that $d''(v)=|C(v)-V_L|$). 
By using a similar technique of updating matching
structures (see \cite{HIOU08}), we can obtain
$\delta((u,v),(*,b))$ in 
$\mO(\Delta^{2/3}(h(v)d''(v))\log^2
\Delta)+\mO(h(v)d''(v))=\mO(\Delta^{2/3}(h(v)d''(v))\log^2 \Delta)$
time. 
Since the number of candidates for $b$ is also bounded by $h(v)$ from 
the neck/head level property, we have the following lemma.  

\medskip

\begin{lemma}\label{maxflow-delta-lem}\rm
$\delta((u,v),(*,*))$ can be computed in $\mO(\Delta^{2/3}(h(v))^2
 d''(v)\log^2 \Delta)$ time, that is, $t(v) = \mO(\Delta^{2/3}(h(v))^2$ $ d''(v)\log^2 \Delta)$. 
\end{lemma}

\medskip

Combining this lemma with 
$\sum_{v \in V-V_L-V_Q}d''(v)\!=\!\mO(n/\Delta)$ 
shown
in Subsection \ref{sec:n175}, we
can show the total running time for the $L(2,1)$-labeling 
 is $\mO(n  
(\max\{h(v)\})^2\cdot$ $(\Delta^{-1/3}\log^2 \Delta))$. 
By applying Theorem \ref{theo:logn}, we have 
the following theorem: 

\medskip

\begin{theorem}\rm 
For trees, the $L(2,1)$-labeling problem can be solved in
$\mO(\min$ $\{n \log^{2}n,\Delta^{1.5}n \})$ time.
Furthermore, if $n=\mO(\Delta^{poly(\log \Delta)})$, it can be solved in
 $\mO(n)$ time. 
\end{theorem}
\begin{corollary}\label{theo:polylog}\rm
For a vertex $v$ in a tree $T$,
we have $\sum_{w \in V(T(v))}t(w)=O(|T(v)|)$
if
  $|T(v)|=\mO(\Delta^{poly(\log \Delta)})$. 
\end{corollary}

\medskip

Only by directly applying Theorem \ref{theo:logn} (actually Lemma
\ref{level-lem}), we obtain much faster running time than the previous
one. In the following section, we present a linear time 
algorithm, in which Lemma \ref{level-lem} is used in a different way.

\section{A Linear Time Algorithm and its Proofs} 
\label{section-linear-running-time}

As mentioned in Subsection \ref{sec:n175},
one of keys for achieving the running time
$\mO(\Delta^{1.5}n)=\mO(n^{1.75})$ is equation $\sum_{v\in V_{\delta}}
d''(v)=\mO(n/\Delta)$, where $V_{\delta}$ is the set of vertices in
which $\delta$-values should be computed via the matching-based algorithm; 
since the computation of $\delta$-values for each
$v$ is done in $\mO(\Delta^{2.5}d''(v))$ time, it takes $\sum_{v\in V_{\delta}}
\mO(\Delta^{2.5}d''(v))=\mO(\Delta^{1.5}n)$ time in total. 
This equation is derived from the fact that in leaf vertices we do not
need to solve the matching to compute $\delta$-values, and any vertex with height 1 has $\Delta
-1$ leaves as its children after the preprocessing operation. 

In our linear time algorithm, we generalize this idea: 
By replacing leaf vertices with subtrees with size at least $\Delta^4$ in
the above argument,
we can obtain  $\sum_{v\in V_{\delta}} d''(v)=\mO(n/\Delta^4)$, and in
total, the running time $\sum_{v\in
V_{\delta}}\mO(\Delta^{2.5}d''(v))=\mO(n)$ is roughly achieved. 
Actually, this argument contains a cheating, because
a subtree with size at most $\Delta^4$
is not always connected to a major vertex, whereas a
leaf is, which is well utilized to obtain $\sum_{v\in V_{\delta}}
d''(v)=\mO(n/\Delta)$. Also, whereas we can neglect leaves to compute 
$\delta$-values, we cannot neglect such subtrees. We resolve these
problems by best utilizing the properties of neck/head levels and
the maximum flow techniques introduced in Section~\ref{section-label-compatibility}.

\subsection{Efficient Assignment of Labels for Computing \boldmath{$\delta$}}

In this section, 
by compiling observations and techniques for
assigning labels in the computation of $\delta((u,v),(*,*))$ for $v \in V$,
given in
Sections~\ref{section-preliminaries} and \ref{section-label-compatibility}, 
\nop{we show that
CK algorithm 
can be implemented to run in linear time.
}
we will design an algorithm to run in linear time within the DP
framework. 
\nop{
In order to implement
 algorithm {\sc CK}  in linear time, 
we 
propose an efficient algorithm for 
computing the value of $\delta((u,v),(*,*))$ for each $v \in V$.}
Throughout this section, we assume that an input tree $T$
satisfies Properties~\ref{leaf-prop} and \ref{P-prop}.
Below, we first partition the vertex set $V$ into  five types of
 subsets defined later,
and give a linear time algorithm for computing the value of $\delta$
 functions,
specified for  each type.

\begin{figure*}[ht]
  \begin{center}
   \input{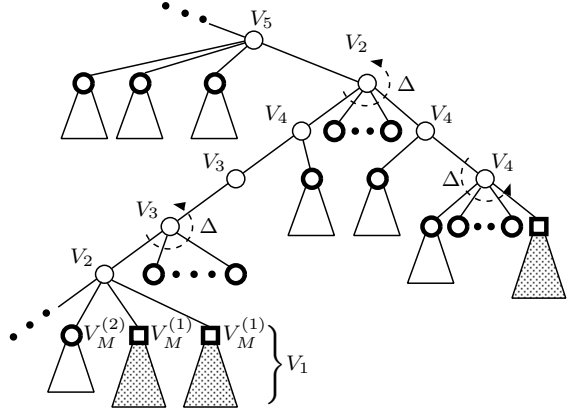}
  \end{center}
 \caption{Partition of $V$ into $V_i$'s $(i=1,\ldots ,5)$. 
Bold circles are leaves $(V_L)$ or pseudo-leaves $(V_M^{(2)}-V_L)$ 
with their subtrees, 
while bold squares are vertices in $V_M^{(1)}$ with their subtrees.}
 \label{fig:v-partition}
\end{figure*}

We here start with defining such five types of  subsets $V_i$ $(i=1,\ldots,5)$.
Let $V_M$ be the set of vertices $v \in V$
such that $T(v)$ is a ``maximal'' subtree of $T$ with $|T(v)|\leq \Delta^5$;
i.e., for the parent $u$ of $v$, $|T(u)|>\Delta^5$.
Divide  $V_M$ into two sets $V_M^{(1)}:=\{v \in V_M \mid 
|T(v)| \geq (\Delta-19)^4 \}$ and  $V_M^{(2)}:=\{v \in V_M \mid 
|T(v)| < (\Delta-19)^4 \}$ (notice that $V_L \subseteq \cup_{v \in V_M}V(T(v))$).
Define $\tilde{d}(v):=|C(v)-V_M^{(2)}|$ $(=d'(v)-|C(v)\cap V_M^{(2)}|)$.
Let
\begin{eqnarray*}
\label{V1-eq}  V_1 &:=& \cup_{v \in V_M}V(T(v)),\\
\label{V2-eq}  V_2 &:=& \{v \in V-V_1 \mid \tilde{d}(v)\geq 2\},\\
\label{V3-eq}  V_3 &:=& \{v \in V-V_1 \mid \tilde{d}(v)=1, 
 C(v)\cap  (V_M^{(2)}- V_L)=\emptyset \},\\
\label{V4-eq}  V_4 &:=& \{v \in V-(V_1\cup V_3) \mid \tilde{d}(v)=1,
 \textstyle{ \sum_{w \in C(v)\cap
 (V_M^{(2)}-V_L)}|T(w)|\leq \Delta(\Delta-19)} \},\\
\label{V5-eq}  V_5 &:=& \{v \in V-(V_1\cup V_3) \mid \tilde{d}(v)=1, 
\textstyle{ \sum_{w \in C(v)\cap
 (V_M^{(2)}-V_L)}|T(w)|> \Delta(\Delta-19)}
 \}.
\end{eqnarray*}
Notice that $V=V_1 \cup V_2 \cup V_3  \cup V_4 \cup V_5$, 
and $V_i \cap V_j=\emptyset$  for each $i,j$ with $i\neq j$ (see Figure
\ref{fig:v-partition}).
In this classification, we call a vertex in $V_M$ a {\em generalized
leaf}, because it plays a similar role of a leaf vertex in
$\mO(n^{1.75})$-time algorithm. 

\nop{
Intuitively, $V_M^{(1)}$ is corresponding to $V_Q$ in the analysis of the 
$\mO(\Delta^{1.5}n)$-time algorithm described in Section 3.
Note that $|V_M^{(1)}| = \mO(n/\Delta^4)$ while $|V_Q| = \mO(n/\Delta)$.
Also, $V_M^{(2)}$ corresponds to $V_L$.
In this sense, each vertex in $V_M^{(2)}-V_L$ can be seen 
as a ``pseudo''-leaf vertex.
Besides, $V_2$ and $V_3$ are corresponding to $V_B$ and $V_P \cup V'_P$, 
respectively.
Each vertex in $V_4 \cup V_5$ is a vertex $v$ which has 
a pseudo-leaf vertex as a child such that if we delete all such
pseudo-leaf vertices from $C(v)$, $v$ becomes a vertex 
in $V_P \cup V'_P$ in the resulting tree.
}


Here we describe an outline of the algorithm for
computing $\delta((u,v),(*,*))$, $v\in V$, named
 {\sc Compute-$\delta$($v$)} (Algorithm \ref{compute-delta}), 
which can be regarded as 
 a subroutine of the DP framework. 
\nop{
\begin{table}[htb]
 \begin{center} 
 \caption{Algorithm {\sc Compute-$\delta$($v$)}}\label{table:CK}
 \begin{tabbing}
  /** Assume that the head and neck levels of $T(v)$ are at most $h$. **/\\ 
 1.\hspace{2pt}
  \=  If $v \in V_1 \cup V_2$, then for each $b \in (L_0-L_h) \cup \{h\}$,
 compute $\delta((u,v),(*,b))$ 
  by the max-flow computation\\
 \>  in 
the network 
 ${\cal N}(u,v,-,b)$ defined in 
  Subsection~\ref{maxflow-sec}.\\
   2. \>  If $v \in V_3$, execute the following procedure
 for each $b \in L_0$ in the case of
  $C(v)\cap V_L =\emptyset$, and for each\\
\>  $b\in \{0,\Delta+1\}$
in the case of $C(v)\cap V_L \neq\emptyset$.
  /** Let $w^*$ denote the unique child of $v$ not in $V_M^{(2)}$.**/\\
2-1. \> \hspace{10pt} \= If $|\{c \mid \delta((v,w^*),(b,c))=1\}|\geq 2$,
  then
 let $\delta((u,v),(*,b)):=1.$\\
2-2. \> \> If  $\{c \mid \delta((v,w^*),(b,c))=1\}= \{c^*\}$, then let
   $\delta((u,v),(c^*,b)):=0$ and
  $\delta((u,v),(a,b)):=1$ for all \\
\> \> other labels $a \notin \{b-1,b,b+1\}$.\\
2-3. \> \> If  $|\{c \mid \delta((v,w^*),(b,c))=1\}|= 0$, then let
   $\delta((u,v),(*,b)):=0$.\\

   3. \> If $v \in V_4\cup V_5$, then similarly to the case of $v \in V_1 \cup V_2$, 
 compute $\delta((u,v),$
  $(*,*))$ 
   by the max-flow \\
\> computation in a network such as ${\cal N}(u,v,-,b)$
  specified for this case (details will be described
 later  \\
\> in Subsections~\ref{V4-sec}  
  and \ref{V5-sec}). 
 \end{tabbing}
 \end{center}
\end{table}
}
%
Below, we show that for each $V_i$,
 $\delta((u,v),(*,*))$, $v \in V_i$ can be computed 
 in linear time in total; i.e., 
$\mO(\sum_{v \in V_i}t(v))=\mO(n)$. 
Namely, we have the following theorem. 
\begin{theorem}\rm
For trees, the $L(2,1)$-labeling problem can be solved in linear time. 
\end{theorem}
\begin{algorithm}[thb]
 \caption{{\sc Compute-$\delta$($v$)}}\label{compute-delta}
 \begin{algorithmic}[1]
\STATE {\small
  /** Assume that the head and neck levels of $T(v)$ are at most
  $h$. **/
\STATE If $v \in V_1 \cup V_2$, then for each $b \in (L_0-L_h) \cup \{h\}$,
 compute $\delta((u,v),(*,b))$ 
  by the max-flow computation
   in the network 
 ${\cal N}(u,v,-,b)$ defined in 
  Subsection~\ref{maxflow-sec}.
\STATE If $v \in V_3$, execute the following procedure
 for each $b \in L_0$ in the case of
  $C(v)\cap V_L =\emptyset$, and for each
  $b\in \{0,\Delta+1\}$
in the case of $C(v)\cap V_L \neq\emptyset$.\\
  /** Let $w^*$ denote the unique child of $v$ not in $V_M^{(2)}$.**/\\
\hspace*{-.5cm}  {\scriptsize 3-1:} \ \ \  If $|\{c \mid \delta((v,w^*),(b,c))=1\}|\geq 2$,
  then
 let $\delta((u,v),(a,b)):=1$ for all labels $a\notin \{b-1,b,b+1\}$,
  and let $\delta((u,v),(a,b)):=0$ for $a\in \{b-1,b,b+1\}$.\\
\hspace*{-.5cm} {\scriptsize 3-2:} \ \ \  If  $\{c \mid
  \delta((v,w^*),(b,c))=1\}= \{c^*\}$, then let
   $\delta((u,v),(a,b)):=0$ for $a \in \{c^*, b-1,b,b+1\}$ and
  $\delta((u,v),(a,b)):=1$ for all 
 the other labels $a \notin \{c^*,b-1,b,b+1\}$.\\
\hspace*{-.5cm} {\scriptsize 3-3:} \ \ \ If  $|\{c \mid \delta((v,w^*),(b,c))=1\}|= 0$, then let
   $\delta((u,v),(*,b)):=0$.
\STATE
    If $v \in V_4\cup V_5$, then similarly to the case of $v \in V_1 \cup V_2$, 
 compute $\delta((u,v),$
  $(*,*))$ 
   by the max-flow 
 computation in a network such as ${\cal N}(u,v,-,b)$
  specified for this case (details will be described
 in Subsections~\ref{V4-sec} and \ref{V5-sec}).}
\end{algorithmic}
\end{algorithm}

We first show $\mO(\sum_{v \in V_1}t(v))=\mO(|V_1|)$.
For each $v \in V_M$, 
$\mO(\sum_{w \in V(T(v))}t(w))$
$=\mO(|T(v)|)$ holds,
by 
Corollary~\ref{theo:polylog} 
and $|T(v)|=\mO(\Delta^5)$.
Hence,
we have
$\mO(\sum_{v \in V_1}t(v))$ $=
\mO(\sum_{v \in V_M}\sum_{w \in V(T(v))}t(w))=$
$\mO(\sum_{v \in V_M}$ $|T(v)|)
=\mO(|V_1|)$.

The proofs for $V_2$, $V_3$, $V_4$ and $V_5$ are given 
in the subsequent subsections. 





\subsection{Computation of \boldmath{$\delta$}-value for \boldmath{$V_2$}}


By Lemma~\ref{maxflow-delta-lem}, it can be seen that 
$\sum_{v \in V_2}t(v)$ 
$ = \mO(\sum_{v \in V_2} \Delta^{2/3}d'(v) h^2 \log^2 \Delta )$
$ = \mO(\Delta^{8/3} \log^2\Delta $ $\sum_{v \in V_2} d'(v)  )$
(note that $h \leq \Delta$ and $d''(v)\le d'(v)$).
Since $d'(v)\leq \Delta$, 
we have 
$
 \sum_{v \in V_2}t(v)= \mO(\Delta^{11/3} \log^2{\Delta}$ $|V_2|).
$
Below, in order to show that $\sum_{v \in V_2}t(v)=\mO(n)$, 
we prove that $|V_2|=\mO(n/\Delta^4)$.

By definition, there is no vertex
whose all children are vertices in $V_M^{(2)}$, since
if there is such a vertex $v$, then  for each $w \in C(v)$,
we have $|T(w)| < (\Delta-19)^4$ and hence $|T(v)| < \Delta^5$, which contradicts
the maximality of $T(w)$.
It follows that
in the tree $T'$ obtained from $T$ by deleting 
all vertices in $V_1-V_M^{(1)}$,
 each leaf vertex belongs to $V_M^{(1)}$
(note that $V(T') = V_M^{(1)} \cup V_2 \cup V_3 \cup V_4\cup V_5$).
Hence, 
\[ \begin{array}{ll}
|V(T')|-1  = |E(T')| &
			 = \frac{1}{2} \sum_{v \in V(T')} d_{T'}(v) \\
			& = \frac{1}{2} (|V_M^{(1)}|+\sum_{v \in V_2
			 \cup V_3 \cup V_4 \cup V_5 }(\tilde{d}(v)+1)-1) \\
			& = \frac{1}{2} (|V_M^{(1)}|+\sum_{v \in V_2}(\tilde{d}(v)+1) + 2|V_3|+2|V_4|+2|V_5|-1) \\
			& \geq
			 \frac{1}{2}|V_M^{(1)}|+\frac{3}{2}|V_2|+|V_3|+|V_4|+|V_5|-\frac{1}{2}
			\end{array} \]
(the last inequality follows from
 $\tilde{d}(v)\geq 2$ for all $v \in V_2$).
Thus, $|V_M^{(1)}|-1 \geq |V_2|$.
It follows by
 $|V_M^{(1)}| = \mO(n/\Delta^4)$
that 
 $|V_2|=\mO(n/\Delta^4)$.



Besides, we can observe that
$
 \sum_{v \in V_2}\tilde{d}(v) = |E(T')|-|V_3|-|V_4|-|V_5| = |V_M^{(1)}|+|V_2|-1
\leq 2|V_M^{(1)}|-2
$
(the first equality follows from 
 $|E(T')|=\sum_{v \in V_2 \cup V_3 \cup V_4 \cup V_5}\tilde{d}(v)$
$=\sum_{v \in V_2}\tilde{d}(v)+|V_3|+|V_4|+|V_5|$
and 
the second equality follows from
 $|E(T')|=|V(T')|-1=|V_M^{(1)}|+|V_2|+|V_3|+|V_4|+|V_5|-1$).
Thus, it holds that 
 $\sum_{v \in V_2} \tilde{d}(v)=\mO(n/\Delta^4)$.
This fact is used in Subsection 4.5.

\subsection{Computation of \boldmath{$\delta$}-value for \boldmath{$V_4$}}
\label{V4-sec}
We first claim that  $|V_4|=\mO(n/\Delta)$. 
Since each leaf is incident to a major vertex
by Property~\ref{leaf-prop},  note that we have 
\begin{equation}\label{>delta-eq}
|T(w)|\geq \Delta, \mbox{ for each $w \in V_M-V_L$}. 
\end{equation}
Now,  we have $C(v)\cap (V_M^{(2)}-V_L)\neq \emptyset$
for $v \in V_4$ by definition.
This implies that 
the total number of descendants
of a child in $V_M^{(2)}-V_L$ of each vertex in $V_4$ is at least 
$\Delta|V_4|$, and hence $|V_4|=\mO(n/\Delta)$.

Here,
 we observe several properties of vertices in $V_4$
before describing an algorithm for   computing $\delta$-values.
Let $v \in V_4$.
By  (\ref{>delta-eq}) and 
$\sum_{w \in C(v)\cap
 (V_M^{(2)}-V_L)}$ $|T(w)|\leq \Delta(\Delta-19)$,
we have
\begin{equation}\label{sizeVM2-eq}
 |C(v)\cap (V_M^{(2)}-V_L)|\leq \Delta-19.
\end{equation}
Intuitively, (\ref{sizeVM2-eq}) indicates that the
number of labels to be assigned for vertices in $C(v) \cap (V_M^{(2)}-V_L)$
is relatively small.
Now,
\begin{equation}\label{level8-eq}
\mbox{the head and neck levels of $T(w)$ are at most 8 for each $w \in
 V_M^{(2)}$}
\end{equation}
by Lemma~\ref{level-lem} and  $|T(w)|<(\Delta-19)^4$ (note that we
assume that $\Delta \ge 18$, since otherwise the original CK algorithm is
already a linear time algorithm). 
Hence,
we can observe that 
there are many possible feasible assignments for
 $C(v) \cap (V_M^{(2)}-V_L)$;
i.e., we can see that if
we can
 assign labels to some restricted children of $v$ properly,
then there exists a proper assignment for the whole children of $v$,
as observed in the following Lemma~\ref{C2-lem}.
For a label $b$,
we divide $C(v)\cap (V_M^{(2)}-V_L)$ into two subsets 
$C_1(b):=\{w \in C(v)\cap (V_M^{(2)}-V_L) \mid \delta((v,w),(b,c))=1$
for all 
$c \in
L_8-\{b-1,b,b+1\}\}$
and
$C_2(b):=\{w \in C(v)\cap (V_M^{(2)}-V_L) \mid \delta((v,w),(b,c))=0$
for all
$c \in L_8-\{b-1,b,b+1\}\}$ (notice that by (\ref{level8-eq}),
the neck level of $T(w)$, $w \in C(v)\cap (V_M^{(2)}-V_L)$,
is at most 8, and hence we have $C(v)\cap (V_M^{(2)}-V_L)=C_1(b) \cup
C_2(b)$).
Let $w^*$ be the unique child of $v$ in $C(v)-V_M^{(2)}$.
By the following property,  we have only to consider the assignments
for $\{w^*\}\cup C_2(b)$.

\nop{
Next consider the case where $v \in V_4$.
For a label $b$,
we divide $C(v)\cap (V_M^{(2)}-V_L)$ into two subsets 
$C_1(b):=\{w \in C(v)\cap (V_M^{(2)}-V_L) \mid \delta((v,w),(b,c))=1$
for all 
$c \in
L_8-\{b-1,b,b+1\}\}$
and
$C_2(b):=\{w \in C(v)\cap (V_M^{(2)}-V_L) \mid \delta((v,w),(b,c))=0$
for all
$c \in L_8-\{b-1,b,b+1\}\}$.
\nop{ (notice that by (\ref{level8-eq}),
the neck level of $T(w)$, $w \in C(v)\cap (V_M^{(2)}-V_L)$,
is at most 8, and hence we have $C(v)\cap (V_M^{(2)}-V_L)=C_1(b) \cup
C_2(b)$).}
By the following property,  we only have to consider the assignments
for $\{w^*\}\cup C_2(b)$.
}

\medskip

\begin{lemma}\label{C2-lem}\rm
For $v \in V_4$, let $a$ and $b$ be labels with $|b-a|\geq 2$
 such that
$b \in L_0$ if $C(v)\cap V_L=\emptyset$
and
$b \in \{0,\Delta+1\}$ otherwise.
Then,  $\delta((u,v),(a,b))=1$ if and only if
 there exists an injective assignment $g: \{w^*\}\cup C_2(b) \rightarrow
 L_0-\{a,b-1,b,b+1\}$ such that $\delta((v,w),(b,g(w)))=1$ for
each $w \in \{w^*\}\cup C_2(b)$. 
\end{lemma}

\medskip

\begin{proof}
 The only if part is clear. We show the if part.
Assume that  there exists an injective assignment
 $g_1: \{w^*\}\cup C_2(b) \rightarrow
 L_0-\{a,b-1,b,b+1\}$
 such that $\delta((v,w),(b,g_1(w)))=1$ for
each $w \in \{w^*\}\cup C_2(b)$.
Notice that by definition of $C_2(b)$,  all $w \in C_2(b)$ satisfies
$g_1(w) \in L_0-L_8$.
Hence, there exists 
at least $|L_8-\{a,b-1,b,b+1,g_1(w^*)\}|\ (=\Delta-19)$ labels which
are not assigned  by $g_1$. By (\ref{sizeVM2-eq}),
we can assign such remaining labels to all vertices in $C_1(b)$
injectively; let $g_2$ be the resulting labeling on $C_1(b)$.
Notice that  for all $w \in
 C_1(b)$, 
  we have $\delta((v,w),(b,g_2(w)))=1$ 
by definition of $C_1(b)$ and $g_2(w) \in L_8$.
It follows that
the function $g_3:C_1(b)\cup C_2(b)\cup \{w^*\}\rightarrow L_0-\{a,b-1,b,b+1\}$ such that
$g_3(w)=g_1(w)$ for all $w \in C_2(b) \cup \{w^*\}$ and
$g_3(w)=g_2(w)$ for all $w \in C_1(b)$ is injective and satisfies
$\delta((v,w),(b,g_3(w)))=1$  for all $w \in C_1(b)\cup C_2(b)$.
Thus, 
if $C(v)\cap V_L=\emptyset$, then
we have $\delta((u,v),(a,b))=1$.

Consider the case where $C(v)\cap V_L\neq \emptyset$.
Let $b=0$ without loss of generality.
Then by Property~\ref{leaf-prop}, $v$ is major.
Hence, $|C(v)\cap V_L|=\Delta-1-|C_1(0)|-|C_2(0)|-|\{w^*\}|$.
Notice that the number of the remaining labels
(i.e., labels not assigned by $g_3$)
is $|L_0-\{0,1,a\}-C_1(0)-C_2(0)-\{w^*\}|=\Delta-2-|C_1(0)|-|C_2(0)|$.
Hence, we can see that
by assigning the remaining labels to vertices in $C(v)\cap V_L$
injectively, we can obtain a proper labeling; 
 $\delta((u,v),(a,b))=1$  holds also in this case.
\end{proof}

Below, we will  show how to compute $\delta((u,v),(*,b))$ 
in $\mO(1)$ time
for a fixed $b$,
where $b \in L_0$ if $C(v)\cap V_L=\emptyset$
and
$b \in \{0,\Delta+1\}$ otherwise.
If $|C_2(b)|\geq 17$, then $\delta((u,v),(*,$ $b))=0$
 because
 in this case, there exists
some $w \in C_2(b)$ to which no label in $L_0-L_8$ can be assigned since $|L_0-L_8|=16$.
Assume that $|C_2(b)|\leq 16$.
There are the following three possible cases:
(Case-1) 
$\delta((v,w^*),(b,c_i))=1$
for  at least two labels $c_1,c_2 \in L_8$,
(Case-2) 
$\delta((v,w^*),(b,$ $c_1))=1$ 
for exactly one label $c_1 \in L_8$,
and (Case-3) otherwise.

(Case-1) By assumption,  for any $a$, $\delta((v,w^*),(b,c))=1$ for some
$c \in L_8-\{a\}$. By Lemma~\ref{C2-lem},
 we only have to check whether there exists
an injective assignment $g:  C_2(b) \rightarrow
 L_0-L_8-\{a,b-1,b,b+1\}$ such that $\delta((v,w),(b,g(w)))=1$ for
each $w \in C_2(b)$.
According to Subsection~\ref{maxflow-sec},
this can be done by utilizing the maximum flow  computation on
 the subgraph ${\cal N}'$ of  ${\cal N}(u,v,-,b)$
induced by $\{s,t\}\cup C_2(b) \cup X'$
where 
 $X'=\{0,1,\ldots,7,$
$\Delta-6,\Delta-5,\ldots,\Delta+1\}$. 
Obviously, the size of ${\cal N}'$ is $\mO(1)$ and
it follows that its time complexity is $\mO(1)$.

(Case-2) For all $a\neq c_1$, the value of 
$\delta((u,v),(a,b))$ can be computed similarly to Case-1.
Consider the case where $a=c_1$. In this case,
if $\delta((v,w^*),(b,c))=1$ holds, then it turns out that $c\in L_0-L_8$.
Hence, by Lemma~\ref{C2-lem},
it suffices to check whether there exists
an injective assignment $g: \{w^*\} \cup  C_2(b)  \rightarrow
 L_0-L_8-\{b-1,b,b+1\}$ such that $\delta((v,w),(b,g(w)))=1$ for
each $w \in \{w^*\} \cup C_2(b)$.
Similarly to Case-1, this can be done in $\mO(1)$ time,
by utilizing  the subgraph ${\cal N}''$ of  ${\cal N}(u,v,-,b)$
induced by $\{s,t\}\cup (C_2(b)\cup \{w^*\}) \cup X'$.

(Case-3) By assumption, 
if $\delta((v,w^*),(b,c))=1$ holds, then it turns out that $c\in L_0-L_8$.
Similarly to the case of $a=c_1$ in Case-2, by using ${\cal N}''$,
we can compute the values of $\delta((u,v),(*,b))$ in $\mO(1)$ time.

We analyze the time complexity for computing $\delta((u,v),$
$(*,*))$.
It is dominated by that for computing $C_1(b)$,
 $C_2(b)$,  and  $\delta((u,v),(*,b))$ for each $b \in L_0$.
By (\ref{level8-eq}), we have $C_i(b)=C_i(b')$ for all $b,b'\in L_8$ and
$i=1,2$.
It follows that the computation of $C_1(b)$ and $C_2(b)$, $b \in L_0$
can be done in $\mO(|C(v)\cap (V_M^{(2)}-V_L)|)$ time.
On the other hand,  the values of $\delta((u,v),(*,b))$ can be computed
in constant time in each case of Cases-1, 2 and 3 for a fixed $b$.
Thus, 
 $\delta((u,v),(*,*))$ can be computed in $\mO(\Delta)$ time.

\subsection{Computation of \boldmath{$\delta$}-value for \boldmath{$V_5$}}
\label{V5-sec}
In the case of $V_4$, since the number of labels to be assigned is 
relatively small, we can properly assign labels to 
$C_1(b)$ even after determining labels for $\{w^*\}\cup C_2(b)$; 
we just need to solve the maximum flow of a small network in which
 vertices corresponding to $C_1(b)$ are omitted. 
In the case of $V_5$, however, the number of labels to be assigned is
not small enough, and actually it is sometimes tight. Thus we need to
assign labels carefully, that is, we need to solve the maximum flow
problem in which the network contains vertices corresponding to $C_1(b)$.
To deal with this, we utilize the  inequality
$ |V_5|\le \frac{n}{\Delta(\Delta-19)}=\mO(n/\Delta^2)$, 
which is derived from 
$\sum_{w \in C(v)\cap
 (V_M^{(2)}-V_L)}$ $|T(w)|> \Delta(\Delta-19)$.

\nop{
%
We first claim that  $|V_4|=\mO(n/\Delta)$. 
Since each leaf is incident to a major vertex
by Property~\ref{leaf-prop}, we note that
for each $u \in V_M-V_L$, we have $|T(u)|\geq \Delta$.
Now, by definition, we have $C(v)\cap V_M\neq \emptyset$
and hence $|V_4|=\mO(n/\Delta)$,
which is derived from the fact that the total number of offspring
vertices of $V_4$ is bounded by $n$. 
}

Below, in order to show that 
$\mO(\sum_{v \in V_5}t(v))=\mO(n)$, 
we prove  that the values of $\delta((u,v),(*,b))$ can be computed 
in $\mO(\Delta)$ time for a fixed $b$.
A key is that
 the children  $w \in C(v)\cap V_M^{(2)}$ of $v$
can be classified into $2^{17}\ (=\mO(1))$ types, 
depending on its $\delta$-values 
$(\delta((v,w),(b,i))\mid i \in (L_0-L_8)\cup \{\tilde{c}_8\})$
where $\tilde{c}_8$ is some label in $L_8-\{b-1,b,b+1\}$,
\nop{$(\delta((v,w),(b,0)),\delta((v,w),(b,1)),
\ldots,$
$\delta((v,w),(b,8)),\delta((v,w),(b,\Delta-6)),\ldots,$
$\delta((v,w),(b,\Delta+1)))$,}
since for each such $w$,
the head and neck levels of $T(w)$ are at most 8,
as observed in (\ref{level8-eq}) (i.e.,
$\delta((v,w),(b,c))=\delta((v,w),(b,\tilde{c}_8))$
for any $c \in L_8-\{b-1,b,b+1\}$). 
We denote the characteristic vector $(\delta((v,w),(b,i))\mid i \in (L_0-L_8)\cup \{\tilde{c}_8\})$
by $(x_w)$.
Furthermore, by the following lemma, we can see that 
$\delta((u,v),(*,b))$ can be obtained
by checking $\delta((u,v),(a,b))$ for $\mO(1)$ candidates of $a$,
where we let $w^*$ be the unique child of $v$ not in $V_M^{(2)}$.

\medskip

\begin{lemma}\label{V5-a-lem}\rm
Let $v\in V_5$ and $b\in L_0$. 
If $\delta((u,v),(a_1,b))\neq \delta((u,v),(a_2,b))$
for some $a_1,a_2 \in L_8-\{b-1,b,b+1\}$ $($say, $\delta((u,v),$
$(a_1,b))=1)$, then we have $\delta((v,w^*),(b,a_2))=1$
and   $\delta((v,w^*),(b,a))=0$ for all $a \in L_8-\{a_2,b-1,b,b+1\}$,
and moreover, $\delta((u,v),(a,b))=1$ for all $a \in L_8-\{a_2,b-1,b,b+1\}$.
\end{lemma}
\medskip
\begin{proof}
Let $f$ be a labeling on $T(u,v)$ with $f(u)=a_1$ and $f(v)=b$,
 achieving
 $\delta((u,v),(a_1,b))=1$.
By  $\delta((u,v),(a_2,b))=0$, 
there exists a child $w_1$ of $v$ with $f(w_1)=a_2$,
since otherwise the labeling from $f$ by changing the label
for $u$ from $a_1$ to $a_2$ would be feasible.

Assume for contradiction that $w_1\neq w^*$ (i.e., $w_1 \in V_M^{(2)}$).
Then the neck level of $T(w_1)$ is at most 8 and
we have $\delta((v,w_1),(b,a_1))=\delta((v,w_1),(b,a_2))=1$ by $a_1,a_2 \in L_8$.
This indicates that $\delta((u,v),(a_2,b))=1$ would hold.
Indeed,  the function $g: C(v)\rightarrow L_0-\{a_2,b-1,b,b+1\}$
such that $g(w_1)=a_1$ and $g(w')=f(w')$ for all other children $w'$ of
 $v$, is injective and satisfies
$\delta((v,w),(b,g(w)))=1$ for all $w \in C(v)$.

\nop{
for a labeling $f_1$ on $(v,w)+T(w)$ achieving
$\delta((v,w),(b,a_1))=1$,
the labeling $f_2$ such that
$f_2(u)=a_2$,
$f_2(x)=f_1(x)$ for all $x \in V(T(w))$
and $f_2(x)=f(x)$ for all other vertices is feasible.
}

Hence, we have $w_1=w^*$. Note
that $\delta((v,w^*),(b,a_2))=1$ since $f$ is feasible.
Then, assume for contradiction that  some $a_3 \in L_8-\{a_2,b-1,b,b+1\}$
 satisfies $\delta((v,w^*),(b,a_3))=1$ ($a_3=a_1$ may hold).
Suppose that there exists a child $w_2$ of $v$ with $f(w_2)=a_3$
(other cases can be treated similarly). 
Notice that the neck level of $T(w_2)$ is at most 8 and
$\delta((v,w_2),(b,a_1))=\delta((v,w_2),(b,a_3))=1$.
Then we can see that  $\delta((u,v),(a_2,b))=1$ would hold.
Indeed,  the function $g: C(v)\rightarrow L_0-\{a_2,b-1,b,b+1\}$
such that $g(w^*)=a_3$, $g(w_2)=a_1$
$g(w')=f(w')$ for all other children $w'$ of
 $v$, is injective and satisfies
$\delta((v,w),(b,g(w)))=1$ for all $w \in C(v)$.
Furthermore, similarly to these observations,
we can see that   $\delta((u,v),(a,b))=1$ 
for all $a \in L_8-\{a_2,b-1,b,b+1\}$.
\end{proof}

~

\noindent
This lemma implies the following two facts.

\medskip

\noindent
{\bf Fact 1:~}In the case of  $|\{c \in L_8-\{b-1,b,b+1\} \mid
\delta((v,w^*),(b,c))=1\}|\neq 1$, 
we have $\delta((u,v),(a,b))=\delta((u,v),(a',b))$ for
all $a,a' \in L_8-\{b-1,b,b+1\}$.

\medskip

\noindent
{\bf Fact 2:~}In the case of
 $\{c \in L_8-\{b-1,b,b+1\} \mid \delta((v,w^*),(b,c))=1\}=\{c^*\}$,
we have
$\delta((u,v),(a,b))=\delta((u,v),(a',b))$ for
all $a,a' \in L_8-\{b-1,b,b+1,c^*\}$ (note that otherwise
Lemma~\ref{V5-a-lem} implies that there would exist
some label $a'' \in L_8-\{b-1,b,b+1\}$ other than $c^*$ with
 $\delta((v,w^*),(b,a''))=1$).

\medskip

\noindent
From these facts, we can observe that
in order to obtain $\delta((u,v),(*,b))$,
we only have to check $a \in (L_0-L_8)\cup \{\tilde{c}_8\}$
in the former case, and
 $a \in (L_0-L_8)\cup \{c^*,c'\}$ where $c' \in L_8-\{b-1,b,b+1,c^*\}$
in the latter case.

\nop{
In the case of $V_4$, since the number of labels to be assigned is 
relatively small, we can properly assign labels to 
$C_1(b)$ even after determining labels for $\{w^*\}\cup C_2(b)$; 
we just need to solve the maximum flow of a small network in which
 nodes corresponding to $C_1(b)$ are omitted. 
In the case of $V_5$, however, the number of labels to be assigned is
not small, and actually it is sometimes tight. Thus we need to
assign labels carefully, that is, we need to solve the maximum flow
problem in which the network contains nodes corresponding to $C_1(b)$.
Instead of that, we can utilize the following inequality
\begin{equation}\label{eq:V5}
 |V_5|\le \frac{n}{\Delta(\Delta-19)}=\mO(n/\Delta^2), 
\end{equation}
which is derived from the fact that the total number of offspring
vertices of $V_5$ is bounded by $n$. 
}

\begin{figure*}[ht]
  \begin{center}
   \input{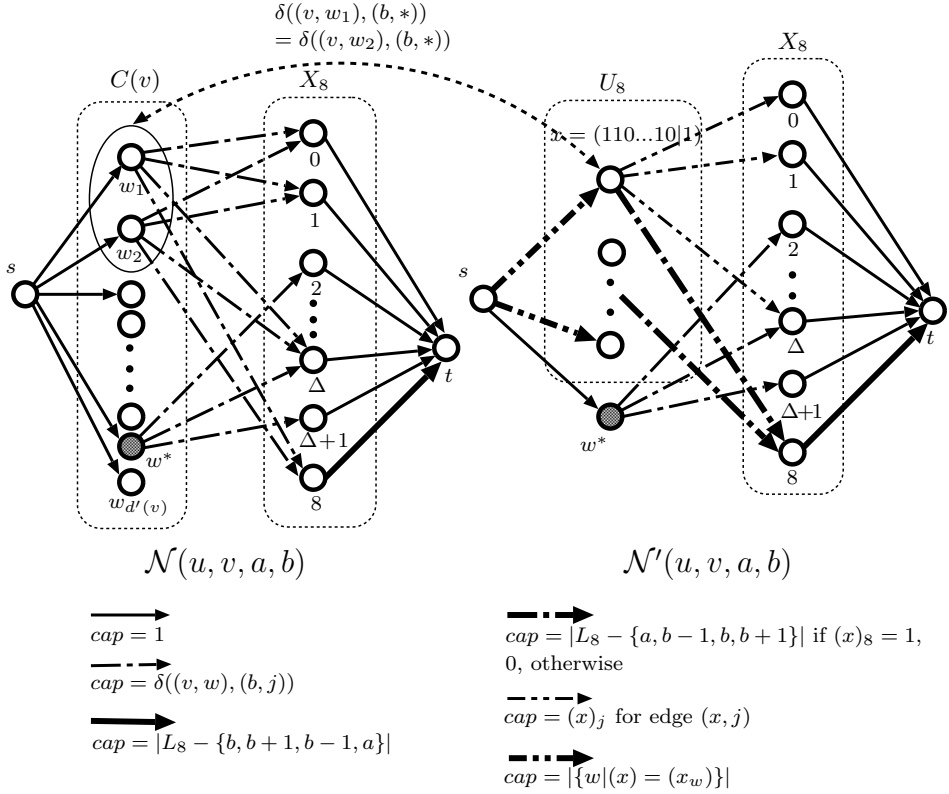}
  \end{center}
 \caption{An example of ${\cal N}(u,v,a,b)$ and ${\cal N'}(u,v,a,b)$ $(h\le 8)$.}
 \label{fig:flow2}
\end{figure*}

From these observations, it suffices to show that
$\delta((u,v),(a,b))$ can be computed in $\mO(\Delta)$ time for a fixed pair $a,b$.
We will prove  this  by applying
  the maximum flow techniques observed 
in Subsection~\ref{maxflow-sec}
to the network  ${\cal
N}'(u,v,a,b)$ with  $\mO(1)$ vertices,
$\mO(1)$ edges, and   $\mO(\Delta)$ units of capacity,
 defined as follows:
\[
 {\cal N}'(u,v,a,b)  = (\{s,t\}\cup U_8 \cup \{w^*\} \cup X_8,  
E_8 \cup E_X' \cup E'_{\delta}, cap),
\]
where
$ U_8  =\{(x_w)\in \{0,1\}^{X_8} \mid w \in C(v)\}$,
$  X_8  =\{0,1, \ldots, 7,
\Delta-6,\Delta-5,\ldots,\Delta+1\} \cup \{8\}$, 
$E_8  =   \{s\} \times (U_8\cup \{w^*\})$,
$E'_{\delta}  = (U_8\cup \{w^*\})\times X_8$, 
$E_X'   =  X_8 \times \{t\}$.
We define $(x_w)_c=\delta((v,w),(b,c))$. 
For example, for $w$, vector $(x_w)=(101\ldots 1|0)$ means that $w$
 satisfies
$(x_w)_0=\delta((v,w),(b,0))=1,(x_w)_1=\delta((v,w),(b,1))=0,(x_w)_2=\delta((v,w),(b,2))=1,\ldots,(x_w)_{\Delta+1}=\delta((v,w),(b,\Delta+1))=1$,
and $(x_w)_{8}=\delta((v,w),$ $(b,8))=0$, where ``$|$'' is put to indicate
the position of $(x_w)_{8}$. 
Notice that  $U_8$ is a set of $0$-$1$ vectors with length 17, and its size is
bounded by a constant. 
When we refer to a vertex $i \in X_8$, we sometimes
use $c_i$ instead of $i$ to avoid a confusion. For 
the edge sets, its
$cap(e)$ function is defined as follows: For $e=(s,(x))\in \{s\}\times
U_8 \subset E_8$, $cap(e)=|\{w \mid (x)=(x_w)\}|$. For $(s,w^*)\in E_8$,
$cap(e)=1$. 
If $a \in X_8- \{c_8\}$, then for $e=(u',a) \in  E_{\delta}'$ with
$u' \in U_8 \cup \{w^*\}$,
$cap(e)=0$.
For $e=((x_w),c)\in U_8\times (X_8- \{a,c_8\}) \subset
E'_{\delta}$, $cap(e)=(x_w)_c$, and for $e=(w^*,c)\in E_{\delta}'$
with $c \in X_8- \{a,c_8\}$,
$cap(e)=\delta((v,w^*),(b,c))$.
For $e=((x_w),c)\in U_8 \times \{c_{8}\} \subset E'_{\delta}$,
 $cap(e)=|L_8-\{a,b-1,b,b+1\}|$ if $(x_w)_8=1$,
 and $cap(e)=0$ otherwise.
For $(w^*, c_{8}) \in E'_{\delta}$, $cap(e)=1$ if
$\delta((v,w^*),(b,c))=1$ for some $c\in L_8 - \{a,b-1,b,b+1\}$, and $cap(e)=0$ otherwise.
For $e=(c,t)\in (X_8 - \{c_8\}) \times \{t\} \subset E_X'$,
$cap(e)=1$. For $e=(c_{8},t)\in E_X'$, $cap(e)=|L_8-\{a,b-1,b,b+1\}|$. 
Figure \ref{fig:flow2} shows an example of relationship between ${\cal
N}(u,v,a,b)$ and ${\cal N}'(u,v,a,b)$, where edges with capacity $0$ are
not drawn. In this example, since
$\delta((v,w_1),(b,*))=\delta((v,w_2),(b,*))$, $w_1$ and $w_2$ 
in the left figure are treated as a single vertex $x=(110\ldots 10|1)$ in the
right figure. 

This network is constructed differently from ${\cal N}(u,v,a,b)$ in 
two
points. One is that in the new ${\cal N}'(u,v,a,b)$, not only 
label vertices but also $C(v)$ vertices are bundled to $U_8$.
For each arc  $e=((x_w),c_8) \in E'_{\delta}$, 
 $cap(e)$ is defined by
$|L_8-\{a,b-1,b,b+1\}|$ if $(x_w)_8=1$
and 0 otherwise.
This follows from the neck level of
$T(w)$ for $w\in C(v)-\{w^*\}$ is at most $8$;
i.e., we have $\delta((v,w),(b,c))=\delta((v,w),(b,c'))$
for all $c,c' \in L_8-\{b-1,b,b+1\}$.
%
The other is that the arc  $(w^*,c_8)$ is set in a different way
from the ones in ${\cal N}(u,v,a,b)$.
Notice that  neck level of
$T(w^*)$ may not be at most 8; $cap((w^*,c_8))=1$ does not imply that
$\delta((v,w^*),(b,c))$ is equal for any $c\in L_8$.
Despite the difference of the definition of $cap$ functions, 
we can  see that 
$\delta((u,v),(a,b))=1$ if and only if there exists a maximum flow 
from $s$ to $t$ with size $d'(v)$ in ${\cal N}'(u,v,a,b)$. 
Indeed, even for a maximum flow $\psi$ with size $d'(v)$ 
such that $\psi(w^*,c_8)=1$ (say, $\delta((v,w^*),(b,c^*))=1$ for
$c^*\in L_8-\{a,b-1,b,b+1\}$), 
there exists an injective assignment $g: C(v)\rightarrow
L_0-\{a,b-1,b,b+1\}$
such that $\delta((v,w),(b,g(w)))=1$ for each $w \in C(v)$,  
since we can assign injectively the remaining labels in $L_8$ (i.e.,
$L_8-\{a,b-1,b,b+1,c^*\}$) to all
 vertices  corresponding to $x_w$ with
$\psi((x_w),c_8)>0$.

To construct ${\cal N}'(u,v,a,b)$, it takes $\mO(d'(v))$ time.  
Since ${\cal N}'(u,v,a,b)$ has $\mO(1)$ vertices, $\mO(1)$ edges
and at most $\Delta$ units of capacity, the maximum flow itself can be solved
in $\mO(\log \Delta)$ time. Thus, the values of
$\delta((u,v),(a,b))$  can be
computed in $\mO(d'(v)+\log \Delta)= \mO(\Delta)$ time.

\subsection{Computation of \boldmath{$\delta$}-value for \boldmath{$V_3$}}


First, we show  the correctness of the procedure in the case of  $v \in V_3$
in algorithm Compute-$\delta$($v$).
Let $v \in V_3$,
 $u$ be the parent of $v$,  $w^*$ be the
unique child of $v$ not in $V_M^{(2)}$, and $b$ be a label
such that $b \in L_0$ if 
$v \in V_3^{(1)}:=\{ v \in V_3 \ |\ C(v) \cap V_L = \emptyset \}$,
 and
$b\in \{0,\Delta+1\}$ if $v \in V_3^{(2)}:=V_3-V_3^{(1)}$.
Notice that 
if $v \in V_3^{(2)}$
(i.e.,  $C(v)\cap V_L \neq\emptyset$), 
 then
by Property~\ref{leaf-prop},
$v$ is major and hence
 $\delta((u,v),(a,b))=1$, $a \in L_0$ indicates
that $b =0$ or $b=\Delta+1$; namely 
we have only to check the case of $b\in \{0,\Delta+1\}$.
Then, if
 there is a label $c \in L_0-\{b-1,b,b+1\}$
such that $\delta((v,w^*),(b,c)) = 1$, then 
 for all $a \in L_0-\{b-1,b,b+1,c\}$,
we have $\delta((u,v),(a,b)) = 1$.
It is not difficult to see that
this observation shows the correctness of the procedure in this case.

Next, we analyze the time complexity of the procedure.
Obviously, for each $v \in V_3$,
we can check   which case of 3-1, 3-2, and
3-3 in algorithm {\sc Compute-$\delta(v)$} holds,
and  determine the values of $\delta((u,v),(*,b))$, in  $\mO(1)$ time.
Therefore, the values of 
 $\delta((u,v),(*,*))$ can be determined in $\mO(\Delta)$ time.
Below,  in order to show that $\sum_{v \in V_3}t(v)=\mO(n)$, 
we prove that $|V_3|=\mO(n/\Delta)$.



As observed above,
each vertex in $v \in V_3^{(2)}$ is major
 and we have
$d(v) = \Delta$. 
Thus, it holds that $|V_3^{(2)}| = \mO(n/\Delta)$.

Finally, we show that $|V_3^{(1)}|=\mO(n/\Delta)$ also holds.
By definition, we can observe that
 for any $v \in V_3^{(1)}$, $d'(v) = \tilde{d}(v) = 1$ (i.e., $d(v)=2$).
By Property~\ref{P-prop}, 
the size of any path component of $T$ is at most 3.
This means that at least $|V_3^{(1)}|/3$ vertices in $V_3^{(1)}$
are children of vertices in $V_2 \cup V_3^{(2)} \cup V_4 \cup V_5$.
Thus, $|V_3^{(1)}|/3 \leq 
\sum_{v \in V_2 \cup V_3^{(2)} \cup V_4 \cup V_5}\tilde{d}(v)$.
From the discussions in the previous subsections (and this subsection),
$\sum_{v \in V_2}\tilde{d}(v) = \mO(n/\Delta^4)$,
 $\sum_{v \in V_3^{(2)}}\tilde{d}(v) = \sum_{v \in V_3^{(2)}}1 = |V_3^{(2)}|= \mO(n/\Delta)$,
 $\sum_{v \in V_4}\tilde{d}(v) = \sum_{v \in V_4}1 = |V_4|= \mO(n/\Delta)$, and
 $\sum_{v \in V_5}\tilde{d}(v) = \sum_{v \in V_5}1 =|V_5|= \mO(n/\Delta^2)$.
Therefore, $|V_3^{(1)}| = \mO(n/\Delta)$.

\section{Extension for \boldmath{$L(p,1)$}-labeling of Trees}\label{sec:lp1}

In the previous sections, we presented an algorithm 
that determines 
whether a given tree $T$ has $(\Delta+1)$-$L(2,1)$-labeling or not 
and showed its linear running time. 
In this section, we extend the result to an algorithm that determines, 
for a given tree $T$ and $\lambda$, 
whether $T$ has $\lambda$-$L(p,1)$-labeling or not. 
For this purpose, we generalize several ideas that contributed 
to the efficient running time of $L(2,1)$-labeling algorithms, 
especially including the notion of label compatibility, 
to be applicable for $L(p,1)$-labeling of trees. 


\smallskip

As seen in the previous sections, 
the linear time $L(2,1)$-labeling algorithm are realized by several
ideas and their combination. 
They are summarized as follows: (1) the equality $\sum_{v\in V_{\delta}} 
d''(v)=\mO(n/\Delta)$ by the preprocessing computation,
(2) label compatibility and the flow-based computation of $\delta$,
which enables a linear time computation for $L(2,1)$-labeling of small
$T$, and
(3) the notion of generalized leaf and classification by the size
of subtrees and the number of generalized leaves, which is also
supported by (1) and (2). 
Namely, if these ideas (especially, (1) and (2))
can be arranged into $\lambda$-$L(p,1)$-labeling, 
the linear time $L(2,1)$-labeling 
algorithm could be extended to a (linear time) algorithm for
$\lambda$-$L(p,1)$-labeling.

In the succeeding subsections, 
we present two properties of $L(p,1)$-labelings 
corresponding to (1) and (2). 
Since it is known that
$\Delta+p-1\le \lambda_{p,1}(T)\le \min\{\Delta+2p-2,2\Delta+p-2\}$,
we assume $\lambda\in [\Delta+p-1,\min\{\Delta+2p-2,2\Delta+p-2\}]$. 
The extended algorithm is based on the same scheme as the linear time
$L(2,1)$-labeling algorithm, that is, it is a dynamic programming
algorithm that computes $\delta$-values from leaves to root. 
In the following, we assume $p$ is a small constant,
unless otherwise stated.  

\subsection{Genralized Major Vertex}\label{section-gmajorvertex}

The extended algorithm also does a similar preprocessing operation to
Algorithm \ref{alg:preprocessing}. 
In the $L(2,1)$-labeling algorithms, Algorithm \ref{alg:preprocessing}
is applied to an input tree so that it satisfies Properties~\ref{leaf-prop}
and~\ref{P-prop}.  
Property~\ref{P-prop} is not affected by the change from
$L(2,1)$ to $L(p,1)$, whereas Property~\ref{leaf-prop} is so. 
The reason why we
can assume Property~\ref{leaf-prop} for an $L(2,1)$-labeling is that 
$T$ has $(\Delta+1)$-$L(2,1)$-labeling if and only if $T-\{v\}$ has
$(\Delta+1)$-$L(2,1)$-labeling, where a leaf vertex $v$ of $T$ 
is adjacent to a non-major vertex. In order to fix this property for 
$\lambda$-$L(p,1)$-labeling, we generalize the notion of major vertex:
A vertex whose degree is $\lambda-p-i+1$ (resp., at least 
$\lambda-p-i+1$) is called {\em $i$-major} (resp., {\em $i^\ge$-major})
with respect to $\lambda$. 
For example, for $\lambda=\Delta+p-1$, a vertex
whose degree is $\Delta-1$ is not $0$-major but $1$-major,
$1^\ge$-major, $2^\ge$-major, and so on, 
with respect to $\Delta+p-1$. We often omit ``with respect to
$\lambda$'', if no confusion arises. 
We can show that $T$ has $\lambda$-$L(p,1)$-labeling 
if and only if $T-\{v\}$ has $\lambda$-$L(p,1)$-labeling where 
leaf $v$ is adjacent to a vertex $u$ whose degree is at most
$\lambda-2p+2$.  
In fact, we can construct $\lambda$-$L(p,1)$-labeling of $T$ from
$\lambda$-$L(p,1)$-labeling of $T-\{v\}$, because in the labeling 
of $T-\{v\}$, $u$ freezes at most $2p-1$ labels and adjacent vertices of 
$u$ freeze at most $\lambda - 2p+1$; $(\lambda+1) -
(2p-1)-(\lambda-2p+1)=1$ label is always available and we can  
use it for labeling $v$ in $T$. Thus we obtain the corresponding version
of Property \ref{leaf-prop}: 

\medskip
\begin{property}\label{leaf-prop2}
 All vertices adjacent to a leaf vertex are $(p-2)^{\ge}$-major vertices
 with respect to $\lambda$. 
\end{property}
\medskip

Consequently, we modify Step 1 of Algorithm
\ref{alg:preprocessing} (i.e., replace ``less than $\Delta$'' with
``less than $\lambda-2p+3$'') so that the resulting $T$ satisfies  
Property \ref{leaf-prop2} instead of Property \ref{leaf-prop}. 

We can see that this change in Algorithm \ref{alg:preprocessing} and 
the generalized notion of $(p-2)^{\ge}$-major vertex absorb several gaps 
caused by the difference between $L(2,1)$ and $L(p,1)$. 
In the analyses of $\mO(n^{1.75})$-time algorithm for $L(2,1)$-labeling 
in Subsection \ref{sec:n175}, for example, 
$\sum_{v\in V-V_L-V_Q}d''(v)$
is estimated as $\mO(n/\Delta)$. By a similar argument, the corresponding
term in the analysis for $L(p,1)$-labeling is estimated as 
$\sum_{v\in
V-V_L-V_Q}d''(v)=\mO(n/(\lambda-2p+3))=\mO(n/(\Delta-p+2))=\mO(n/\Delta)$,
which implies that  
an $\mO(\Delta^{1.5}n)$-time algorithm
for $\lambda$-$L(p,1)$-labeling of trees is obtained 
only by this minor modification.  


\subsection{Genralized Label Compatibility and its Application} 
As
mentioned several times,  
the most important idea in the analyses of the linear time algorithm for
$L(2,1)$-labeling of trees is the label compatibility. 
In this subsection, we generalize it to $\lambda$-$L(p,1)$-labeling of
trees. In the following, we use $\delta_{\lambda}((*,*),(*,*))$ 
to denote the DP-tables in the $\lambda$-$L(p,1)$-labeling algorithm. 

We say that $T$ is $\lambda$-{\em head}-$L_h$-{\em compatible} if
             $\delta_{\lambda}((u,v),(a,b))=\delta_{\lambda}((u,v),(a',b))$ for
             all $a,a'\in L_h$ and $b\in L_0$ with $|a-b|\ge p$ and
             $|a'-b|\ge p$. 
	     Analogously, we say that $T$ is $\lambda$-{\em
             neck}-$L_h$-{\em compatible} if
	     $\delta_{\lambda}((u,v),(a,b)) =
             \delta_{\lambda}((u,v),(a,b'))$             for all $a\in
             L_0$ and $b,b' \in L_h$  
             with $|a-b|\ge p$ and $|a-b'|\ge p$. 
We often omit ``$\lambda$'' if it is clear. 
The neck and head levels of $T$ with respect to $\lambda$ are 
defined similarly to Definition \ref{def:labelcomatibility21}:

\medskip

\begin{definition}\label{def:labelcomatibilityp1}
 Let $T$ be a tree rooted at $v$, and $u\not\in V(T)$.\\
 {\rm (i)}       The {\rm neck level} $($resp., {\rm head level}$)$ of
 $T$ with respect to $\lambda$ is $0$ if $T$ is 
             $\lambda$-neck-$L_0$-compatible $($resp.,
 $\lambda$-head-$L_0$-compatible$)$.  
{\rm (ii)}
             The {\rm neck level} $($resp., {\rm head level}$)$ of $T$
 with respect to $\lambda$ is
             $h\ (\geq 1)$ if $T$ is not
             $\lambda$-neck-$L_{h-1}$-compatible $($resp.,
 $\lambda$-head-$L_{h-1}$-compatible$)$ 
             but $\lambda$-neck-$L_h$-compatible $($resp., $\lambda$-head-$L_h$-compatible$)$. 
\end{definition}

\medskip

\noindent 

For $L(2,1)$-labeling, Lemma \ref{level-lem} gives 
a relationship between the neck/head levels and the tree size. 
This lemma is generalized as follows: 

\medskip  
\begin{lemma}\label{level-lem2}\rm
(Level Lemma (Generalized Form)) 
Let $T'$ be a subtree of $T$.
{\rm (i)} If $|T'|< (\lambda-2h-4p+4)^{h/(2p-2)}$, $\lambda-2h\geq
 3p-3$,
and $p>1$,  then
the head level and neck level of $T'$ 
are both at
 most $h$. 
{\rm (ii)} If $p=1$, then
the head level and neck level of $T'$ are 0.
\nop{If $|V(T')|\leq (\Delta-3-2h)^{h/2}-1$ and $\Delta-2h\geq 10$, then
the head level and neck level of $T'$ are both at most $h$. 
}
\end{lemma}

\medskip  

\noindent 
Note that $p$ is a general positive integer in Lemma
\ref{level-lem2}.
By letting $\lambda=\Delta+1$ and $p=2$ in Lemma \ref{level-lem2},
we immediately obtain Lemma \ref{level-lem}. 
Similarly to Theorem \ref{theo:logn}, 
we obtain the following theorem: 

\medskip  

\begin{theorem}\rm\label{theo:logn2}
 For a tree $T$ and a positive integer $p$, both the
 head and neck levels of $T$ with respect to $\lambda$ are
$\mO(\min\{\Delta,p\log |T|/\log \lambda\})$.
 \end{theorem}

\begin{corollary}\rm\label{cor:logn2}
If $p$ is bounded by a constant and $\lambda\in
[\Delta+p-1,\Delta+2p-2]$, the head and neck levels of $T$ with respect
 to $\lambda$ are $\mO(\min\{\Delta,\log |T|/\log \Delta\})$.
\end{corollary}
 
\medskip  

The proofs of Lemma \ref{level-lem2} and Theorem \ref{theo:logn2} 
are shown in Section \ref{sec:proof-level-lem}. 

\medskip  

Similarly to Subsection \ref{maxflow-sec}, this theorem enables 
to speed up flow-based computation of $\delta$.
The running time of calculating 
$\delta_{\lambda}((u,v),(*,b))$  for a pair $(a,b)$ 
is
\[
 \mO((h(v)+d''(v))^{2/3}(h(v)d''(v))\log
(h(v)+d''(v))\log \Delta)
= \mO(\Delta^{2/3}(h(v)d''(v))\log^2 \Delta),  
\]
where $h$ and $d''$ are defined similarly 
as in Subsection \ref{maxflow-sec}.   
We thus have the following lemma.  

\medskip

\begin{lemma}\rm\label{maxflow-delta-lem2}
$\delta_{\lambda}((u,v),(*,*))$ can be computed in
 $\mO(\Delta^{2/3}(h(v))^2  d''(v)\log^2 \Delta)$ time for a positive integer $p$.
\end{lemma}

\medskip

Since we assume $p$ is bounded by a constant, we have $\sum_{v\in
V-V_L-V_Q}d''(v)=\mO(n/\Delta)$ (as we saw in the previous subsection) and
$h(v)=\mO(\min\{\Delta,\log |T|/\log \Delta\})$
(Corollary \ref{cor:logn2}). 
By these, the following theorem is obtained: 

\medskip

\begin{theorem}\rm 
 Given a tree $T$ and positive integers $\lambda$ and $p$, 
it can be determined in
$\mO(\min\{ \Delta^{2/3}$ $p^2 n\log^{2}n,\Delta^{2.5}n\})$ 
time if $T$ has $\lambda$-$L(p,1)$-labeling.
 If $p$ is a positive constant integer, it can be 
done in $\mO(\min\{n\log^{2}n,\Delta^{1.5}n\})$ time. 
Furthermore, if $n=\mO(\Delta^{poly(\log \Delta)})$, it can be determined in
 $\mO(n)$ time. 
\end{theorem}

\bigskip

As seen in these subsections, 
the crucial properties in the analyses for that $L(2,1)$-labeling 
of trees is solvable in linear time are generalized 
to $\lambda$-$L(p,1)$-labeling if $p$ is bounded by a positive constant.
This implies that all the analyses in Section 
\ref{section-linear-running-time} are extended straightforwardly to the
case of a constant $p$, which yields a linear time
$\lambda$-$L(p,1)$-labeling algorithm for a constant $p$, 
though we omit the details to avoid tedious repetitions. 
Since the $L(p,1)$-labeling number of a tree $T$ is in 
$[\Delta+p-1,\min\{\Delta+2p-2,2\Delta+p-2\}]$, 
an optimal $L(p,1)$-labeling of $T$ can be obtained 
by applying the $\lambda$-$L(p,1)$-labeling algorithm for $T$ at most
$p$ times; the total running time for solving $L(p,1)$-labeling problem
for $T$ is $\mO(n)$ if $p$ is a constant. 

\medskip

\begin{theorem}\rm 
For trees, the $L(p,1)$-labeling problem can be solved in linear time,
 if $p$ is bounded by a positive constant. 
\end{theorem}

\section{Proof of Level Lemma}\label{sec:proof-level-lem}
In this section, we provide proofs of Lemma \ref{level-lem2} and Theorem
\ref{theo:logn2}. 
For a tree $T'$ rooted at $v$, denote by $T'+(u,v)$ the
tree obtained from $T'$ by 
adding a  vertex $u \notin V(T')$ and an edge $(u,v)$.
This is  similar to
$T(u,v)$ defined in Subsection~\ref{CK-algo-sec},
however,
for $T(u,v)$,
 $u$ is
regarded as a virtual vertex, while for $T'+(u,v)$,
$u$ may be an existing vertex.

~

\nop{
\begin{lemma}\rm
Let $T'$ be a subtree of $T$.
If $|T'|\leq (\Delta-3-2h)^{h/2}-1$ and $\Delta-2h\geq 10$, then
the head level and neck level of $T'$ are both at most $h$. 
\end{lemma}
}

\noindent
{\bf Proof of Lemma~\ref{level-lem2}:}
%
(i) Let $p>1$,  $\delta((u,v),(*,*)):=\delta_{\lambda}((u,v),(*,*))$,
and $h>0$ (note that the case of $h=0$ means $|T'|=0$).
When $h\leq 2p-3$, we have $|T'|\leq \lambda-4p+1$
 and hence
$\Delta(T'+(u,v))\leq \lambda-4p+1$, where
$v$ denotes the root of $T'$.
It follows that  $T'+(u,v)$ can be labeled by using at most 
 $\Delta(T'+(u,v))+2p-1\leq \lambda-2p$
  labels.
Thus, in these cases, it is not difficult to see that 
 the head 
and neck levels are both 0.

Now, we assume for contradiction that this lemma does not hold.
Let $T_1$ be such a counterexample with the minimum size, i.e.,
$T_1$ satisfies the following properties (\ref{level1-eq})--(\ref{level4-eq}):

\vspace*{.1cm}

\noindent
\begin{eqnarray}
\label{level1-eq} && \ |T_1|< (\lambda-2h-4p+4)^{h/(2p-2)}, \\
\label{level2-eq} && \mbox{ the neck or head level of $T_1$ is
 at least $h+1$,}\\
\label{level3-eq} && \mbox{ $h \geq 2p-2$ (from the arguments of the previous paragraph), and } \\
\label{level4-eq} && \mbox{ for each tree $T'$ with $|T'|<|T_1|$, the lemma holds.}
\end{eqnarray}

\vspace*{.1cm}

\noindent
By (\ref{level2-eq}), there are two possible cases (Case-I)
the head level of $T_1$ is at least $h+1$ and
(Case-II) the neck level of $T_1$ is at least  $h+1$.
Let $v_1$ denote the root  of $T_1$.

(Case-I) In this case, in $T_1+(u,v_1)$ with $u \notin V(T_1)$,
for some label $b$,
there exist two labels $a,a' \in L_h$ with $|b-a|\geq p$ and $|b-a'|\geq p$
such that $\delta((u,v_1),(a,b))=1$ and $\delta((u,v_1),(a',b))=0$.
Let $f$ be a $\lambda$-$L(p,1)$-labeling with $f(u)=a$ and $f(v_1)=b$.
If any child of $v_1$ does not have label $a'$ in the labeling $f$, 
then the labeling
obtained from $f$ by changing the label for $u$ from $a$
to $a'$
is also feasible, which contradicts  $\delta((u,v_1),(a',b))=0$.

Consider the case where some child $w$ of $v_1$ satisfies $f(w)=a'$.
Then by $|T(w)|<|T_1|$ and
 $(\ref{level4-eq})$, the neck level of $T(w)$ is at most
 $h$.
Hence, we have $\delta((v_1,w),(b,a'))=$
 $\delta((v_1,w),(b,a))\ (=1)$ by $a,a' \in L_h$.
Let $f_1$ be a $\lambda$-$L(p,1)$-labeling
on $T(w)+(v_1,w)$ achieving  $\delta((v_1,w),(b,a))=1$. 
Now, note that any vertex $v \in C(v_1)-\{w\}$ satisfies
$f(v) \notin \{a,a'\}$, since $f$ is feasible.
Thus, we can observe that the labeling $f_2$ satisfying
the following (i)--(iii) is  a $\lambda$-$L(p,1)$-labeling on $T_1+(u,v_1)$:
(i) $f_2(u):=a'$,
(ii) $f_2(v):=f_1(v)$ for all vertices $v \in V(T(w))$,
and (iii) $f_2(v):=f(v)$ for all other vertices.
This also contradicts  $\delta((u,v_1),(a',b))=0$.

(Case-II) By the above arguments, we can assume that
the head level of $T_1$ is at most $h$.
In $T_1+(u,v_1)$ with $u \notin V(T_1)$,
for some label $a$,
there exist two labels $b,b' \in L_h$ with $|b-a|\geq p$ and $|b'-a|\geq p$
such that $\delta((u,v_1),(a,b))=1$ and $\delta((u,v_1),(a,b'))=0$.
Similarly to Case-I, we will derive a contradiction  
by showing that $\delta((u,v_1),(a,b'))=1$.
Now there are the following three cases:
(II-1) there exists such a pair $b,b'$ with $b'=b-1$,
(II-2) the case (II-1) does not hold and there exists such a pair $b,b'$ with $b'=b+1$,
and (II-3) otherwise.

First we show that we have only to consider the case of (II-1);
namely, we can see that if  
\begin{eqnarray}\label{assumption-eq}
 & & \mbox{(II-1) does not occur for any }a, b \mbox{ with }|b-a|\geq p,
  |b-1-a|\geq p, \\ \nonumber & & \mbox{ and }
  \{b-1,b\}\subseteq L_h,
\end{eqnarray}
 then
neither (II-2) nor (II-3) occurs. 
Assume that
(\ref{assumption-eq}) holds.
Consider the case (II-2). Then, since we have 
$\delta((u,v_1),(\lambda-a,\lambda-b))=\delta((u,v_1),(a,b))$ $=1$
and 
$\delta((u,v_1),(\lambda-a,\lambda-b-1))=\delta((u,v_1),(a,b+1))=0$,
it contradicts  (\ref{assumption-eq}).
Consider the case (II-3), that is, 
there is no pair $b,b'$ such that $|b-b'|=1$.
Namely, in this case, 
for some $a \in L_{h+p}$, 
we have $\delta((u,v_1),(a,h))\neq \delta((u,v_1),(a,\lambda-h))$
and
 $\delta((u,v_1),(a,b_1))=\delta((u,v_1),(a,b_2))$
only if  (i) $b_i\leq a-p$, $i=1,2$ or (ii) $b_i \geq a+p$, $i=1,2$.
Then, 
since the head level of $T_1$ is at most $h$, 
 it follows by $a \in L_{h+p}$ that $\delta((u,v_1),(a,h))=
\delta((u,v_1),(\lambda-a,h))$.
By $\delta((u,v_1),(\lambda-a,h))=\delta((u,v_1),(a,\lambda-h))$,
we have  $\delta((u,v_1),(a,h))= \delta((u,v_1),(a,\lambda-h))$,
a contradiction.

Below, in order to show (\ref{assumption-eq}),
we consider the case of $b'=b-1$.
Let $f$ be a $\lambda$-$L(p,1)$-labeling with $f(u)=a$ and $f(v_1)=b$.
We first  start with the labeling $f$,
and change the label
for $v_1$
from $b$ to $b-1$.
Let $f_1$ denote the resulting labeling.
If $f_1$ is feasible, then it contradicts $\delta((u,v_1),(a,b-1))=0$.
Here, we assume that $f_1$ is infeasible,
and will show how to construct another $\lambda$-$L(p,1)$-labeling
by changing the assignments for vertices in $V(T_1)-\{v_1\}$.
Notice that since $f_1$ is infeasible, there are
\begin{eqnarray}
\label{child-eq} &&\mbox{some child $w$ of $v_1$ with $f_1(w)=b-p$, or} \\
\label{gchild-eq}&&\mbox{some grandchild $x$ of $v_1$ with
 $f_1(x)=b-1$.} 
\end{eqnarray}
Now we have the following claim.

\begin{claim}\label{level1-cl}
Let $f'$ be a $\lambda$-$L(p,1)$-labeling on $T_1$
and
$T(v)$ be a subtree of $T_1$.
There are at most
$\lambda-2h-4p+3$ children $w$ of $v$ with
$f'(w)\in L_h$ and
$|T(w)|\geq (\lambda-2h)^{(h-2p+2)/(2p-2)}$.
\nop{
 one of the following
 properties {\rm (i)--(iii)} hold:\\
$(i)$
 $|L'|\geq 2$. \\
$(ii)$ 
 $|L'|=1$ and
there exists a child $w$ of $v$ with
$f'(w) \in L_h$ and $|T(w)|\leq (\Delta-2h)^{(h-2)/2}-1$.\\
$(iii)$ 
There exist two children $w_1,w_2$ of $v$ with
$\{f'(w_1),f'(w_2)\} \subseteq L_h$,
 $|T(w_1)|\leq (\Delta-2h)^{(h-2)/2}-1$,
and
$|T(w_2)|\leq (\Delta-2h-2)^{(h-1)/2}-1$.}
\end{claim}
\begin{proof}
Let $C'(v)$ be the set of children $w$ of $v$ with
$f'(w)\in L_h$ and $|T(w)|\geq (\lambda-2h)^{(h-2p+2)/(2p-2)}$.
If this claim does not hold, then
we would have $|T_1|\geq |T(v)|\geq 1+ \sum_{w \in C'(v)}|T(w)|$
$\geq 1 + (\lambda-2h-4p+4)(\lambda-2h)^{(h-2p+2)/(2p-2)}> 1+ (\lambda-2h-4p+4)^{h/(2p-2)}>
 |T_1|$,
a contradiction. 
\end{proof}

\vspace*{.2cm}

\noindent
This claim indicates that given a feasible labeling $f'$
of $T_1$,
for each vertex $v \in V(T_1)$,
there exist at least $2p-2$ labels $\ell_1,\ell_2,\ldots,\ell_{2p-2} \in L_h$
such that $\ell_i$ is not assigned to any vertex in $\{v,p(v)\}\cup
 C(v)$ (i.e., $\ell_i \notin \{f'(v') \mid v' \in \{v,p(v)\}\cup C(v)\}$)
 or assigned to a child $c_i \in C(v)$  with $|T(c_i)|<
 (\lambda-2h)^{(h-2p+2)/(2p-2)}$, since
 $|L_h-\{f'(p(v)),f'(v)-p+1,f'(v)-p,\ldots,f'(v)+p-1\}|\geq
\lambda-2h-2p+1$,
where $p(v)$ denotes  the parent of $v$.
For each vertex $v \in V(T_1)$,
denote such labels by $\ell_i(v;f')$  and
such children by $c_i(v;f')$ (if exists)
for $i=1,2,\ldots,2p-2$.
We note that by (\ref{level4-eq}), if $c_i(v;f')$ exists, then
the head  and neck levels of $T(c_i(v;f'))$ are at most $h-2p+2$.

First consider  the case where the vertex of (\ref{child-eq}) exists;
denote such a vertex by $w_1$.
We consider this case by dividing into  two cases (II-1-1)
$b\geq h+p$ and (II-1-2) $b\leq h+p-1$.

(II-1-1) Suppose that  we have $\ell_1(v_1;f)\neq b-p$,
and $c_1(v_1;f)$   exists
 (other cases can be
treated similarly). 
By (\ref{level4-eq}), the head  and neck levels of
$T(w)$ are at most $h$ for each $w \in C(v_1)$, and especially,
the head  and neck levels of
$T(c_1(v_1;f))$ are at most $h-2p+2$.
Hence, we have
\begin{eqnarray}
\nonumber\delta((v_1,w_1),(b,b-p))&=&\delta((v_1,w_1),(b,\ell_1(v_1;f)))\\
\label{6-eq}&= &\delta((v_1,w_1),(b-1,\ell_1(v_1;f))),\\
\nonumber\delta((v_1,c_1(v_1;f)),(b,\ell_1(v_1;f)))&=&\delta((v_1,c_1(v_1;f)),(b-1,\ell_1(v_1;f)))\\
\label{7-eq}&=&\delta((v_1,c_1(v_1;f)),(b-1,b+p-1)), 
\end{eqnarray}
since $\ell_1(v_1;f)\notin \{b-p,b-p+1,\ldots,b+p-1\}$ and
 $\{b-p,b-1,b,\ell_1(v_1;f)\}\subseteq L_h$
(note that in the case where $c_1(v_1;f)$ does not exist,
 (\ref{7-eq}) is not necessary).
Notice that  $b+p-1\notin L_h$ may hold, however
we have $b+p-1 \in L_{h-p+1}$ by $b \in L_h$.
By these observations,
there exist labelings $f_1'$ and $f_2'$ 
of  $T(w_1)+(v_1,w_1)$ and $T(c_1(v_1;f))+(v_1,c_1(v_1;f))$,
achieving 
$\delta((v_1,w_1),(b-1,\ell_1(v_1;f)))=1$
and 
$\delta((v_1,c_1(v_1;f)),(b-1,b+p-1))=1$,
respectively. 
Let $f^*$ be the labeling
such that $f^*(v_1)=b-1$,
$f^*(v)=f_1'(v)$ for all $v \in V(T(w_1))$,
$f^*(v)=f_2'(v)$ for all $v \in V(T(c_1(v_1;f)))$,
and
$f^*(v)=f(v)$ for all other vertices.

(II-1-2) In this case, we have $b+2p-2 \in L_h$ 
 by $\lambda-2h \geq 3p-3$.
Now we have the following claim.
\begin{claim}\label{2-2-cl}
For $T(w_1)+(v_1,w_1)$, we have $\delta((v_1,w_1),(b-1,b+p-1))=1$. 
\end{claim}
\begin{proof}
Let $f_1$ be the labeling such that $f_1(v_1):=b-1$,
$f_1(w_1):=b+p-1$, and $f_1(v):=f(v)$ for all other vertices $v$.
Assume that $f_1$ is
infeasible to $T(w_1)+(v_1,w_1)$
since otherwise the claim is proved.
Hence, (A) there exists some child $x$ of $w_1$ with
 $f_1(x)\in \{b+1,b+2,\ldots,b+2p-2\}$
or (B) some grandchild $y$ of $w_1$ with 
$f_1(y)=b+p-1$ 
(note that
 any child $x'$ of $w_1$ has neither label $b-1$ nor $b$ by 
$f(w_1)=b-p$ and
$f(v_1)=b$).

First, we consider the case where vertices of (A) exist.
Suppose that  there are $2p-2$ children $x_1,\ldots,x_{2p-2} \in C(w_1)$
with $f_1(x_i)=b+i$, 
 we have $\{\ell_i(w_1;f)\mid i=1,2,\ldots,2p-2\}\cap \{f(x_i) \mid i=1,2,\ldots,2p-2\}=\emptyset$,
and all of $c_i(w_1;f)$   exist
 (other cases can be
treated similarly).
\nop{Let
 $b'' \in L_h- \{h,h+1,\ldots,b+p-3\} - (\cup_{i=1}^{2p-2}\{\ell_i(w_1)-p+1,\ell_i(w_1)-p+2,\ldots,\ell_i(w_1)+p-1\})$ (such $b''$ exists by
$\lambda-2h\geq 4p(p-1)-1$). }

Now by (\ref{level4-eq}), the head and neck levels of
$T(x_i)$ (resp., $T(c_i(w_1;f))$) is at most $h$ (resp., $h-2p+2$) for $i=1,2,\ldots,2p-2$.
Hence, for each $i=1,2,\ldots,2p-2$, we have
\begin{eqnarray}
\label{6'-eq}\delta((w_1,x_i),(b-p,b+i))&=&\delta((w_1,x_i),(b-p,\ell_i(w_1;f)))\\
\nonumber\delta((w_1,c_i(w_1;f)),(b-p,\ell_i(w_1;f)))&=&\delta((w_1,c_i(w_1;f)),(\lambda-h+p,\ell_i(w_1;f)))\\
\nonumber&=&\delta((w_1,c_i(w_1;f)),(\lambda-h+p,b-2p+i)) \\
\label{7'-eq}&=&\delta((w_1,c_i(w_1;f)),(b+p-1,b-2p+i)), 
\nop{\\
\label{8'-eq}\delta((w_1,x_2),(h-1,h+3))&=&\delta((w_1,x_2),(h-1,\ell_2(w_1;f)))\\
\nonumber\delta((w_1,c_2(w_1;f)),(h-1,\ell_2(w_1;f)))&=&\delta((w_1,c_2(w_1;f)),(b'',\ell_2(w_1;f)))\\
\nonumber&=&\delta((w_1,c_2(w_1;f)),(b'',h-1)) \\
\label{9'-eq}&=&\delta((w_1,c_2(w_1;f)),(h+2,h-1)),
}
\end{eqnarray}
since  we have $\{b+i,\ell_i(w_1;f)
\} \subseteq L_h$,
$|(b-p)-\ell_i(w_1;f)|\geq p$, 
$|(\lambda-h+p)-(b-2p+i)|\geq p$,
$|(\lambda-h+p)-\ell_i(w_1;f)|\geq p$,
and
$b-2p+i \in L_{h-2p+2}$ (by $b\geq h+1$)
for
 $i=1,2,\ldots,2p-2$,
and $\{b+p-1,\lambda-h+p\}\subseteq L_{h-2p+2}$ (notice that $b+p-1\in
 L_h$
also hold).
Also notice that $b-2p+i\geq h-2p+2\geq 0$ by $b\geq h+1$ and
 (\ref{level3-eq}).
By (\ref{6'-eq}) and (\ref{7'-eq}), 
there exist $\lambda$-$L(p,1)$-labelings
 $f_i'$ and $f_i''$, $i=1,2,\ldots,2p-2$, 
of $T(x_i)+(w_1,x_i)$, $T(c_i(w_1;f))+(w_1,c_i(w_1;f))$, 
achieving 
$\delta((w_1,x_i),(b-p,\ell_i(w_1;f)))=1$ and
$\delta((w_1,$ $c_i(w_1;f)),(b+p-1,b-2p+i))=1$,
\nop{$\delta((w_1,x_2),(h-1,\ell_2(w_1;f))))=1$,
and
$\delta((w_1,$ $c_2(w_1;f)),(h+2,h-1))=1$,}
respectively.
Let $f_2$ be the  labeling of $T(w_1)+(v_1,w_1)$ such that
$f_2(v_1)=b-1$, 
$f_2(w_1)=b+p-1$,
$f_2(v)=f_i'(v)$ for all   $v \in V(T(x_i))$, 
$f_2(v)=f_i''(v)$ for all   $v \in V(T(c_i(w_1;f)))$, 
and 
$f_2(v)=f(v)$ for all other vertices.
Observe that
we have 
$f_2(x_i)=\ell_i(w_1;f)$,
$f_2(c_i(w_1;f))=b-2p+i$, 
 and
$f_2(x)\notin \{b-1, b,\ldots,b+2p-2\}$ for all $x \in C(w_1)$, 
every two labels in $C(w_1)$ are pairwise disjoint, and
$f_2$ is a $\lambda$-$L(p,1)$-labeling of each subtree $T(x)$ with
$x \in C(w_1)$.

Assume that $f_2$ is still infeasible. Then, there 
 exists some grandchild $y$ of $w_1$ with $f_2(y)=b+p-1$.
Observe that 
from $f(w_1)=b-p$,
no sibling of 
such a grandchild $y$ has label $b-p$ in the labeling $f_2$,
 while  such $y$ may exist in the subtree $T(x)$
with $x \in C(w_1)-
  \{c_i(w_1;f)\mid i=1,2,\ldots,2p-2\}$.
Also note that 
for the parent $x_p=p(y)$ of such $y$,
we have $f(x_p)\notin \{b-2p+1,\ldots,b-1\}$.
Suppose that 
 $\ell_1(x_p;f_2)\neq b+p-1$ holds and $c_1(x_p;f_2)$ exists
(other cases can be treated similarly).
Now by (\ref{level4-eq}), the neck level of $T(y)$ (resp., 
$T(c_1(x_p;f_2))$) is at most $h$ (resp., $h-2p+2$).
Hence, we have 
$\delta((x_p,y),(f_2(x_p),b+p-1))=$
 $\delta((x_p,y),(f_2(x_p),\ell_1(x_p;f_2)))$
and 
$\delta((x_p,c_1(x_p;f_2)),(f_2(x_p),\ell_1(x_p;f_2)))$
 $=\delta((x_p,c_1(x_p;f_2)),(f_2(x_p),$ $b-p)))$.
It follows that there
exist $\lambda$-$L(p,1)$-labelings
$f''_1$ and $f''_2$ on 
$T(y)+(x_p,y)$ and $T(c_1(x_p;f_2))+(x_p,c_1(x_p;f_2))$
which achieve
 $\delta((x_p,y),(f_2(x_p),$ $\ell_1(x_p;f_2)))=1$
and
 $\delta((x_p,c_1(x_p;f_2))$, $(f_2(x_p),$ $b-p)))=1$,
respectively.
It is not difficult to see that
the labeling $f''$ such that 
$f''(v)=f''_1(v)$ for all $v \in V(T(y))$,
$f''(v)=f''_2(v)$ for all $v \in V(T(c_1(x_p;f_2)))$,
and $f''(v)=f_2(v)$ for all other vertices
is 
a $\lambda$-$L(p,1)$-labeling of $T(x_p)+(w_1,x_p)$.

Thus, by repeating these observations for each grandchild $y$ of $w_1$ with
$f_2(y)=b+p-1$, we can obtain a 
 $\lambda$-$L(p,1)$-labeling $f_3$ of $T(w_1)+(v_1,w_1)$ with
$f_3(v_1)=b-1$ and $f_3(w_1)=b+p-1$.
\end{proof}

\medskip

\noindent
Let $f^*$ be the labeling such that
$f^*(v)=f_3(v)$ for all  $v \in \{v_1\}\cup V(T(w_1))$
and $f^*(v)=f(v)$ for all other vertices.
Thus, in both cases (II-1-1) and (II-1-2), we have constructed
a labeling $f^*$ such that $f^*(u)=a$, $f^*(v_1)=b-1$, and
$f^*(w)\notin \{a, b-p,b-p+1,\ldots,b+p-2\}$ for all $w \in C(v_1)$, 
every two labels in $C(v_1)$ are pairwise disjoint, and
$f^*$ is a $\lambda$-$L(p,1)$-labeling on each subtree $T(w)$ with
$w \in C(v_1)$.

Assume that $f^*$ is still infeasible.
Then, there exists  some grandchild $x$ of $v_1$ of (\ref{gchild-eq}).
Notice that for each  vertex $v \in \{p(x)\} \cup V(T(x))$,
we have $f^*(v)=f(v)$ from the construction; $f^*(p(x))\notin
 \{b-p+1,\ldots,b+p-1\}$
and $f^*(x')\neq b$ for any sibling $x'$ of $x$.
Moreover, by (\ref{level4-eq}), the neck level of $T(x)$ is at most $h$;
$\delta((p(x),x),(f^*(p(x)),b-1))$
$=\delta((p(x),x),(f^*(p(x)),b))=1$.
Hence,
there exists
a $\lambda$-$L(p,1)$-labeling $f'$ of $T(x)+(p(x),x)$
which achieves $\delta((p(x),x),(f^*(p(x)),b))=1$.
It follows that  the labeling $f''$
such that $f''(v)=f'(v)$ for all $v \in V(T(x))$
and $f''(v)=f^*(v)$ for all other vertices, is
 a $\lambda$-$L(p,1)$-labeling of $T(p(x))+(v_1,p(x))$.
Thus, by repeating these observations for each grandchild of $v_1$ of
 (\ref{gchild-eq}), we can obtain a 
 $\lambda$-$L(p,1)$-labeling $f^{**}$ for $T_1+(u,v_1)$ with
$f^{**}(u)=a$ and $f^{**}(v_1)=b-1$.
This contradicts $\delta((u,v_1),(a,b-1))=0$. 

(ii) 
When $p=1$, the constraint $|a-b|\ge 1$ for any labels $a$ and $b$ is 
equivalent to $a\neq b$; in any $L(1,1)$-labeling, we can exchange any
two different labels. This implies (ii). 
\qed

\medskip 

As mentioned before, this lemma holds for general $p$. 
so does Theorem \ref{theo:logn2}, and to show this, notice 
that the following easy lemma holds. 

\medskip

\begin{lemma}\label{p>=2delta:lem}\rm
If $|T|\geq 2$ and $p \ge 2\Delta$, no label in $\{2\Delta-1,\ldots, p-1\}$ 
is used in any
 $\lambda$-$L(p,1)$-labeling of a tree $T$ for any $\lambda \in \{\Delta+p-1,\ldots,2\Delta+p-2\}$. 
\end{lemma}

\medskip

\noindent
{\bf Proof of Theorem~\ref{theo:logn2}:}
We first show that the head level and neck level of $T$ is
 $\mO(\Delta)$.
The case of $p=\mO(\Delta)$ is clear because
$\lambda(T)\leq \Delta+p-2$.
If $p \geq 2\Delta$, then Lemma~\ref{p>=2delta:lem}
indicates that the head level and neck level of $T$ is at most
$2\Delta-2$ (note that if $|T|=1$, both levels are 0).

We next show that the head level and neck level of $T$ is
 $\mO(p\log{n}/\log{\lambda})$. 
The case of $\lambda=\mO(p\log{n}/\log{\lambda})$
 is clear.
 Consider the case where $\lambda> \frac{8p\log{n}}{\log{(\lambda/2)}}+8p-8$.
Then, for $h=\frac{2p\log{n}}{\log{(\lambda/2)}}$,
we have 
\begin{eqnarray}
 \nonumber  (\lambda-2h-4p+4)^{\frac{h}{(2p-2)}} & > &
  \left(\frac{\lambda}{2}+\left(\frac{4p\log{n}}{\log{\frac{\lambda}{2}}}+4p-4\right)-\frac{4p\log{n}}{\log{\frac{\lambda}{2}}}-4p+4\right)^{\frac{\log{n}}{\log{\frac{\lambda}{2}}}}\\
\nonumber & = & n.
\end{eqnarray}
Now note that  $\lambda-2h>\frac{4p\log{n}}{\log{(\lambda/2)}}+8p-8>3p-3$.
Hence, by Lemma~\ref{level-lem2}, it follows that
the head  and neck levels of $T$ are both at most 
$\frac{2p\log{n}}{\log{(\lambda/2)}}$.
\qed


\section{Concluding Remarks}\label{sec:conclusion}

This paper presents a linear time algorithm for $L(p,1)$-labeling of
trees, when $p$ is bounded by a constant, 
especially by describing the one in case of $p=2$, 
which is the $L(2,1)$-labeling problem. 
Although the main contribution of the paper is the linear time algorithm
itself, the introduction of the notion of label-compatibility might have
more impact, because it could be generalized for other distance
constrained labelings.   



\end{document}